\documentclass[aps, ,prb,longbibliography,floatfix,superscriptaddress,twocolumn,showpacs,amsmath,amssymb,reprint,nofootinbib]{revtex4-2}
\usepackage[normalem]{ulem}
\usepackage{amsmath}
\usepackage{bigints}
\usepackage{amsfonts}
\usepackage{amssymb}
\usepackage{bm}
\usepackage{adjustbox}
\usepackage{epstopdf}
\usepackage{color}
\usepackage{textcomp}
\usepackage{gensymb}
\usepackage{graphicx}
\usepackage[dvipsnames]{xcolor}

\usepackage{dcolumn}
\usepackage{bm}
\usepackage{braket}
\usepackage{hyperref}
\usepackage{booktabs}

\usepackage{verbatim}
\usepackage{physics}

\usepackage{hyperref}
\hypersetup{
    colorlinks=true,
    linkcolor=BurntOrange,
    urlcolor=BrickRed,
    citecolor=Mulberry,
    pdftitle={Theory for Cd3As2 thin films in the presence of magnetic fields},
    pdfauthor={M. Smith, V. L. Quito, A. A. Burkov, P. P. Orth, I. Martin}
}

\definecolor{orange}{RGB}{255,127,0}

\begin{document}

\title{Theory for Cd$_3$As$_2$ thin films in the presence of magnetic fields}
\author{M. Smith}
\affiliation{Materials Science Division, Argonne National Laboratory, Lemont, Illinois 60439, USA}
\author{V. L. Quito}
\affiliation{Ames National Laboratory, Ames, Iowa 50011, USA}
\affiliation{Department of Physics and Astronomy, Iowa State University, Ames, Iowa 50011, USA}
\affiliation{S\~{a}o Carlos Institute of Physics, University of S\~{a}o Paulo,
PO Box 369, 13560-970, S\~{a}o Carlos, SP, Brazil.}
\author{A.A. Burkov}
\affiliation{Department of Physics and Astronomy, University of Waterloo, Waterloo, Ontario N2L 3G1, Canada}
\affiliation{Perimeter Institute for Theoretical Physics, Waterloo, Ontario N2L 2Y5, Canada}
\author{P.~P. Orth}
\affiliation{Ames National Laboratory, Ames, Iowa 50011, USA}
\affiliation{Department of Physics and Astronomy, Iowa State University, Ames, Iowa 50011, USA}
\affiliation{Department of Physics, Saarland University, 66123 Saarbr\"ucken, Germany}
\author{I. Martin}
\affiliation{Materials Science Division, Argonne National Laboratory, Lemont, Illinois 60439, USA}
\date{\today}

\begin{abstract}
   We present a theory for thin films of the Dirac semimetal  Cd$_3$As$_2$  in the presence of magnetic fields. We show that, above a critical thickness, specific subbands $n$ of thin film Cd$_3$As$_2$ are in a quantum spin Hall insulator regime and study their response to in- and out-of-plane magnetic fields. We find that sufficiently large in-plane Zeeman fields drive the system toward a 2D Dirac semimetal regime, provided the field is directed perpendicular to a high-symmetry mirror plane. For other directions, we find the Dirac points to be weakly gapped. We further investigate how the system responds to finite out-of-plane field components, both starting from the quantum spin Hall regime at small in-plane fields and from the 2D Dirac semimetal regimes at larger in-plane fields, addressing recent experimental observations in [A.~C. Lygo \emph{et al.}, Phys. Rev. Lett. {\bf 130} 046201 (2023)] and [B.~Guo \emph{et al.}, Phys. Rev. Lett. {\bf 131}, 046601(2023)]. 
\end{abstract}

\maketitle

\addtolength{\abovedisplayskip}{-1mm}

\section{Introduction \label{sec:Introduction}}
The materials search for quantum spin Hall (QSH) insulators has been a major topic in condensed matter physics ever since their theoretical proposal ~\cite{KaneMele_PRL_2005,BernevigHughesZhang_Science_2006,BernevigZhang_PRL_2006,HasanKane_RMP_2010, maciejkoQuantumSpinHall2011, qianQuantumSpinHall2014,CaoChen_AdvQuanTech_2019,WangWan_PRB_2019}. Due to their intriguing and robust transport properties, which arise from topologically protected helical edge states and can be controlled by electric and magnetic fields, these materials are interesting both from a fundamental perspective as well as for technological applications, e.g., for a topological field-effect transistor~\cite{feiEdgeConductionMonolayer2017}. The QSH effect has thus far been realized in various systems such as semiconductor quantum wells~\cite{KonigZhang_Science_2007, RothZhang_Science_2009,DuDu_PRL_2015}, heavy-element analogs of graphene such as bismuthene~\cite{reisBismutheneSiCSubstrate2017} and germanene~\cite{bampoulisQuantumSpinHall2023}, and transition metal dichalcogenide monolayers~\cite{feiEdgeConductionMonolayer2017, Tang_NatPhys_2017, WuJarillo_Herrero_Science_2018}. 
Another promising route to realizing QSH states is to start from three-dimensional (3D) bulk topological insulators (TIs) or semimetals (TSMs) and attempt to reach the QSH regime by tuning quantum confinement in thin films or heterostructures~\cite{linderAnomalousFiniteSize2009, lu_massive_2010,liu_oscillatory_2010,ZyuzinBurkov_PRB_2011}. 
In TIs, the very thin film limit is characterized by a hybridization between the top and the bottom 2D surface states, creating a gapped low energy spectrum~\cite{BernevigHughesZhang_Science_2006,zhangCrossoverThreedimensionalTopological2010,
Burkov_PRB_2018,Akzyanov_PRB_2019,ChorsiSchuller_AdvOpts_2020}. In contrast, in a TSM with surface normal directed along the axis that separates the bulk nodes, there exists no top and bottom surface states to hybridize, and instead, the bulk spectrum is gapped due to quantum confinement~\cite{liu_oscillatory_2010,lu_massive_2010}.
This situation has recently been experimentally realized in thin films of Cd$_3$As$_2$~\cite{Stemmer_PRLa_2023,Stemmer_PRLb_2023}, which is a prototypical 3D Dirac TSM in the bulk~\cite{wang_three-dimensional_2013}. It was driven to the thin film limit in epitaxial $\approx 20$ nm thick films that were grown along the $[001]$ crystallographic direction, which is parallel to the separation direction between the bulk Dirac nodes. Various transport properties were measured in the presence of magnetic fields~\cite{Stemmer_PRLa_2023,Stemmer_PRLb_2023}; explaining these is a primary motivation for our work. 
Thus, here we theoretically consider Cd$_3$As$_2$ thin films that terminate along the [001] crystallographic direction and investigate their band structure and edge states in the presence of a magnetic field. We characterize their dependence on film thickness, chemical potential, and magnetic field strength and direction, a crucial step to understanding Cd$_3$As$_2$ thin film properties.

In thin films of Cd$_3$As$_2$ the transverse momentum quantization leads to a reorganization of the 3D Dirac dispersion into 2D subbands. The subbands are generally gapped, although the gaps can be small, making the effects of external magnetic fields potentially large. Recent magnetotransport measurements in a perpendicular magnetic field along the $[001]$ direction~\cite{Stemmer_PRLa_2023} revealed a magnetic field dependence of the Landau level spectrum that is in agreement with that of a 2D TI~\cite{KonigZhang_Science_2007}. Subsequent experiments in tilted magnetic fields with both in- and out-of-plane components showed an odd-integer quantum Hall effect~\cite{Stemmer_PRLb_2023}. This was attributed to a transition to a 2D Dirac semimetal phase, with the two Dirac cones appearing along momentum space directions perpendicular to the in-plane magnetic field. 

Motivated by these recent experimental findings, we develop and analyze a model for the low-energy electronic states in $[001]$ thin films in the presence of magnetic fields pointing along general directions. Specifically, we construct a low-energy $k \cdot p$ band structure model that considers the contributions of the different types of electrons close to the Fermi level. We demonstrate that the $s$ and relevant $p$ orbitals experience in-plane magnetic fields in a very distinct way: due to the strong spin-orbit coupling, the relevant $p$ orbitals show a cubic Zeeman effect, while the $s$ orbitals experience the standard linear Zeeman coupling.
This distinct coupling manifestly conserves the $C_{4z}$ symmetry of the system and differentiates our model from recent work in Ref.~\cite{MiaoDai_arxiv_2023}, which considers a linear coupling of magnetic field to the relevant $p$ orbitals. We show below that the two models have distinct behaviors as a function of in-plane field orientation. 
Nevertheless, both models similarly yield a Dirac semimetal phase for fields pointed in the $[100]$ direction, which Ref.~\cite{MiaoDai_arxiv_2023} refers to as a 2D Weyl semimetal. Ref.~\cite{MiaoDai_arxiv_2023} does not consider the effect of rotations of the in-plane field and focuses on non-trivially gapping out the 2D Dirac nodes via the application of lattice strain. Here, we show that the cubic $p$-orbital Zeeman effect for in-plane magnetic fields leads to a 2D Dirac semimetal phase only when the field is applied in a direction perpendicular to a mirror plane and naturally leads to a gapped spectrum for fields rotated away from these high symmetry directions. 

Specifically, we derive the proper coupling of the relevant $p$ orbitals to the magnetic field, showing it is consistent with all symmetry operations, which are explicitly given in Appendix~\ref{sec:App_Adding-a-magnetic-field}. We then investigate the band structure as a function of in-plane magnetic field. For specific field directions (perpendicular to a mirror plane), we find that there exists a critical field strength such that the system transitions from a gapped to a gapless band structure with two Dirac points appearing along the line perpendicular to the applied field (\emph{i.e.}, along the mirror line). As the field is rotated away from these high symmetry directions, the Dirac points become gapped. Interestingly, applying an out-of-plane Zeeman field can close and reopen these Dirac gaps, leading to a change of the Chern number in the system. We further use this model to explore the orbital effect of the perpendicular component of the magnetic field when the system is driven into the 2D Dirac semimetal regime by an in-plane field. We find there is an emergent quantum number for the low-lying Landau levels, the Dirac states, which causes an additional two-fold Landau level degeneracy. These findings provide a plausible explanation for the recent quantum Hall measurements reported in Ref.~\cite{Stemmer_PRLb_2023}. Lastly, we discuss the evolution of the surface states in the presence of an in-plane Zeeman field, showing that the conventional edge states are coupled by the in-plane magnetic field and thus gapped and eventually disappear for large enough in-plane field strength.

The remainder of this paper is structured as follows: Section~\ref{sec:Thin-film} establishes the effective low-energy theory for Cd$_3$As$_2$ thin films with surface normal along the [001] direction. The resulting model is generally applicable to Dirac TSMs with a surface normal that is parallel to the direction separating the Dirac points. Section~\ref{sec:In-plane-B} discusses the electronic response to in-plane Zeeman fields, first only considering the Zeeman coupling of the $s-$orbitals and then deriving the smaller (nonlinear) Zeeman effect in the relevant $p-$orbitals.
Section~\ref{sec:Surface-states} discusses the surface states of the thin film with and without in-plane Zeeman coupling.
Section~\ref{sec:Out-plane-B} considers the effect of an out-of-plane orbital magnetic field. We first discuss a purely out-of-plane field and then include a possibly large in-plane field component. 
Finally, in Section~\ref{sec:Conclusions}, we summarize our findings and predictions.

\section{Model for (001) thin films}
 \label{sec:Thin-film}
Here, we derive a low-energy model of Cd$_3$As$_2$ thin films, grown such that the surface normal points along the $[001]$ direction. We start from a low-energy four-band model for bulk Cd$_3$As$_2$. The low-energy bands are located around the $\Gamma$ point in the Brillouin zone and include the $\ket{S_{1/2}, m_J=\pm 1/2}$ and $\ket{P_{3/2},m_J = \pm\ 3/2}$ states~\cite{wang_three-dimensional_2013}. The relevant point group constraining the model is $D_{4h}$ (or $4/mmm$), and up to cubic order in momentum, it reads
\begin{align}
H_{0}&(\mathbf{k}) =\epsilon_0(\mathbf{k}) \tau_{0} s_{0}+ \mathcal M(\mathbf{k}) \tau_{z} s_{0}\nonumber \\
 & +A\left(k_{x} \tau_{x} s_{z}-k_{y} \tau_{y} s_{0}\right)+C_{3a}k_{x}k_{y}\left(k_{x}\tau_{y} s_{0}-k_{y}\tau_{x} s_{z}\right)\nonumber \\
 & +C_{3b}\left(k_{x}^{3}\tau_{x} s_{z}- k_{y}^{3}\tau_{y} s_{0} \right)+C_{3c}k_{z}^{2}\left(k_{y} \tau_{y} s_{0}- k_{x}\tau_{x} s_{z}\right) \nonumber\\
 &+C_{3d} (k_x^2 - k_y^2)k_z \tau_x s_x   - C_{3e} k_x k_y k_z \tau_x s_y   \,.
\label{eq:H0-third-order}
\end{align}
Here, $s_i$ and $\tau_i$ are Pauli matrices in angular momentum ($m_J = \pm$) and orbital ($S, P$) space, respectively, and we have defined 
\begin{subequations}
\begin{align}
	\epsilon_0(\mathbf{k}) &= C_0 + C_1 k_z^2 + C_2 k_{\parallel}^2\\
	\mathcal M(\mathbf{k}) &= M_0 - M_z k_z^2  - M_{xy} k_{\parallel}^2
\end{align}
\end{subequations}
with $k_{\parallel}^2 = k_x^2 + k_y^2$ and $M_{xy}, M_z>0$, consistent with~\cite{Olle_PRB_2020}. The $k_z$ direction points along the fourfold rotation axis, corresponding to the $[001]$ direction. 
A detailed derivation of this effective model is provided in Appendix~\ref{sec:App_kp_model}. Up to quadratic order, it agrees with the Hamiltonian derived in Refs.~\cite{wang_three-dimensional_2013,Olle_PRB_2020,Stemmer_PRLa_2023}. The four-dimensional basis is obtained by projecting the two $s$ and six $p$ bands onto the total angular momentum basis $J$ and keeping the low energy states $\ket{S_{1/2}, m_J=\pm 1/2}$ and $\ket{P_{3/2},m_J = \pm\ 3/2}$. The four other $p$-orbital states lie approximately $250$ meV higher in energy~\cite{wang_three-dimensional_2013}. 
The low-energy model for a film 
of thickness $L$ is constructed by imposing open boundary conditions along the $[001]$ direction. The Hamiltonian at zero in-plane momentum, $\mathbf{k}_{\parallel} = 0$, reads
\begin{align}
	H_{0}|_{\mathbf{k}_{\parallel} = 0} = (C_0 - C_1 \partial_z^2) + (M_0 + M_{z} \partial_z^2)s_0 \tau_z 
 \label{eq:H0_k_parallel_equal_zero}
\end{align}
Let us focus on the $m_J = +$ subspace spanned only by the $\ket{S_{1/2},1/2}$ and $\ket{P_{3/2}, 3/2}$  states as the energy levels of the opposite angular momentum, $m_J= -$, are degenerate. Since $H_{\mathbf{k}_\parallel = 0}$ is diagonal, the orbital structure of the eigenstates read $\psi_+ = (1,0)^T$ and $\psi_- = (0,1)^T$, \emph{i.e.} they localized on the $S$ and $P$ states, respectively. Using the trial wavefunctions $\Psi_{\beta n}(z) = A \psi_\beta e^{\lambda_{\beta n} z}$, and imposing boundary conditions such that the wavefunctions vanish at $z = \pm L/2$, we find the wavefunction for $l = 0, 1, 2, \ldots$ and $\beta = \pm$ is given by
\begin{align}
	\Psi_{\beta n} (z) &=
	\sqrt{\frac{2}{L}} \psi_\beta \begin{cases}
		\sin(2l\pi \frac{z}{L}), \quad n= 2l\\
		\cos( (2l+1) \pi\frac{z}{L} ), \quad n=2l+1 \,.
	\end{cases} \label{eq:WF_kzero}
\end{align}
The associated energies at $\mathbf{k}_\parallel = 0$ read
\begin{align}
	E_{\beta n} &= C_0 + \beta M_0 + \left(\frac{\pi n}{L}\right)^2 \left(C_1 - \beta M_{z} \right).
\end{align}
Here $n = 1, 2, 3, \ldots$ is the sub-band index, and $\beta = \pm$ denotes the orbital index. Thus, the subbands for $[001]$ thin films are found by setting $k_z = \pi n/L$. The Hamiltonian for a specific subband takes the form of Eq.~\eqref{eq:H0-third-order} with
\begin{subequations}
    \begin{align}
        \epsilon_{0} \left(\mathbf{k}\right) & \rightarrow\epsilon_{n}\left(\mathbf{k}_\parallel\right)=C_{0}+C_{1}\left(\frac{\pi n}{L}\right)^{2}+C_{2} k_{\parallel}^{2} 
        \label{eq:eps_n}
        \\
        \mathcal M\left(\mathbf{k}\right) & \rightarrow \mathcal M_{n}\left(\mathbf{k}_\parallel\right)=M_{0}-M_{z}\left(\frac{\pi n}{L}\right)^{2}-M_{xy} k_{\parallel}^{2}\,.
        \label{eq:matcal_n}
    \end{align}
    \label{eq:epsilon_and_M_for_subband_n}
\end{subequations}
The total Hamiltonian thus involves 4 different bands for each subband $n$. In the following, we use the notation $\mathbf{k}\equiv\mathbf{k}_{\parallel}$ unless specified otherwise.
Importantly, the topological character of the bands is determined by
\begin{equation}
    \mathcal M_n(0) \equiv M_n = M_0 - M_{z}\left(\frac{n \pi}{L}\right)^2
    \label{eq:mass_subband_n} \,.
\end{equation}
As $M_{xy} > 0$, for positive $M_n >0$ the bands are inverted and topologically non-trivial, while for negative $M_n < 0 $, the bands un-invert and become topologically trivial.
Considering the film thicknesses reported in Refs.~\cite{Stemmer_PRLa_2023, Stemmer_PRLb_2023}, we find that for a thickness $L \sim 19$~nm it is the $n=2$ subband which is most relevant. It exhibits a topological gap $M_{n=2} \sim 5.5$ meV and the $n=1,3$ subbands are $\sim 20$ meV away. The relatively small topological gap of the $n=2$ subband and the large effective $g$-factor of the Zeeman coupling make thin film Cd$_3$As$_2$ easily tunable by in-plane Zeeman fields. As will be shown below, the addition of an in-plane Zeeman field causes the topological gap for the $m_J = \pm$ states to be different, leading to a gapless regime for specific Zeeman field directions.

\section{In-plane Magnetic Field Effects \label{sec:In-plane-B}}
In this Section, we show the effects of adding in-plane magnetic fields $\mathbf{B} = B (\cos \phi, \sin \phi, 0)$.  We start with a derivation of the Zeeman coupling to a magnetic field in the $xy$ plane.

\subsection{Microscopic derivation of the Hamiltonian}
The magnetic field $\mathbf{B}$ couples to the $s$ and $p$ states according to the microscopic Zeeman Hamiltonian
\begin{equation}
H_{Z} = g_{s}\mu_B\mathbf{B}\cdot\left(\mathbf{L}+2\mathbf{S}\right)+g_{p}\mu_B\mathbf{B}\cdot\left(\mathbf{L}+2\mathbf{S}\right). \label{eq:Zeeman_field}
\end{equation}
Here, $\mathbf{L}$ and $\mathbf{S}$ are the orbital and spin angular momentum operators, $\mu_B$ is the Bohr magneton, and $g_s, g_p$ are the $g$-factors associated with $s$ and $p$ states. 
Given the large spin-orbit coupling in the system, it is convenient to work with the total angular momentum $J$. The states close to the Fermi surface are the doublets of $J=3/2$, $\ket{P_{3/2}, \pm \frac{3}{2}}$, and of $J=1/2$, $\ket{S_{1/2}, \pm 1/2}$~\cite{wang_three-dimensional_2013}. The $s$ electrons of the Kramers pair $\ket{S_{1/2}, \pm 1/2}$ couple to the in-plane magnetic field as $H_{Z}^{(S)} = g_s \mu_B \mathbf{B}\cdot \mathbf{s}$, where $\mathbf{s}= (s_x, s_y, s_z)$ is the vector of the Pauli matrices. 
The low energy $p$ states $\ket{P_{3/2}, \pm 3/2}$, however, do not couple linearly to the in-plane component of the field. Instead, the in-plane magnetic field couples the $\ket{P_{3/2}, \pm \frac{3}{2}}$ only in the  cubic order. This can be understood as follows: terms in the Hamiltonian linear in the applied in-plane field can only change the azimuthal quantum number $m_J$ by $1$, and therefore, the components $m_J=\pm3/2$ are not coupled to linear order in the field; cubic coupling is, however, possible, involving transition via $\ket{P_{1/2}, \pm \frac{1}{2}}$ and $\ket{P_{3/2},\pm \frac{1}{2}}$ virtual states. The explicit calculations deriving this coupling are performed in Appendix~\ref{sec:App_Adding-a-magnetic-field}. 

To lowest order in an in-plane magnetic field $\mathbf{B} \equiv \mathbf{B}_\parallel = (B_x, B_y, 0)$, the effective coupling of the $\ket{P_{3/2}, \pm \frac{3}{2}}$ states reads
\begin{align}
H_{Z}^{(P)}(\mathbf{B}_\parallel) =\frac{g^3_{p}\mu_B^3}{\Delta^{2}}&\Bigl[B_{x}\bigl(B_{x}^{2} -3B_{y}^{2}\bigr)s_{x} \nonumber \\ & \quad  -B_{y}\bigl(B_{y}^{2}-3B_{x}^{2}\bigr)s_{y}\Bigr] \,.
\label{eq:p-coupled-B-par}   
\end{align}
Here, $\Delta$ is the energy scale of the higher bands at the $\Gamma$ point (see the derivation in Appendix~\ref{sec:App_Adding-a-magnetic-field}). 
Thus, the in-plane Zeeman field terms for the effective low-energy model are given by 
\begin{align}
H(&\mathbf{B}_\parallel) =g_{s}\mu_B\left(\frac{\tau_{0}+\tau_{z}}{2}\right)\mathbf{B}_{\parallel}\cdot\mathbf{s}+\frac{g_{p}^3\mu_B^3}{\Delta^{2}}\left(\frac{\tau_{0}-\tau_{z}}{2}\right) \nonumber \\
 & \times \left[B_{x}\left(B_{x}^{2}-3B_{y}^{2}\right)s_{x}-B_{y}\left(B_{y}^{2}-3B_{x}^{2}\right)s_{y}\right]. \label{eq:HB_parallel}
\end{align}
It is convenient to parameterize the field in the $xy$ plane as $B_{x}=B_\parallel \cos \phi $ and $B_{y}=B_\parallel \sin \phi $. Returning to Eq.~\eqref{eq:H0-third-order}, we simplify our calculation by choosing coefficients $C_{3a} = 3 C_{3b} = 3 A_1$,$C_{3c} = 0$,  and $C_{3e} = -C_{3d}/2 = -2\eta$, which simplifies Eq.~\eqref{eq:H0-third-order} to its atomic limit form (see Appendix~\ref{sec:App_Arb_Coeff} for further details). 
Finally, the full Hamiltonian, including the in-plane field, is found from Eq.~\eqref{eq:H0-third-order} combined with Eq~\eqref{eq:HB_parallel}. Under the assumption that the finite-$k$ terms do not lead to significant changes in the wavefunctions along the $z$ direction, we project the Hamiltonian into states of Eq.~\eqref{eq:WF_kzero}, leading to the effective Hamiltonian

\begin{align}
     H &= \epsilon_n(\mathbf{k}) \tau_0s_0 + \mathcal M_n(\mathbf{k}) \tau_z s_0 + A(k_x \tau_x s_z - k_y \tau_y s_0) \nonumber\\
    & +3 A_1 k_{x}k_{y}\left(k_{x}\tau_{y} s_{0} -k_{y}\tau_{x} s_{z} \right) + A_1 \left(k_{x}^{3}\tau_{x} s_{z}-k_{y}^{3}\tau_{y}s_{0} \right) \nonumber\\
    &+ \eta \frac{n\pi}{L} \left[ (k_x^2 - k_y^2) \tau_x s_x + 2 k_x k_y \tau_x s_y\right] \nonumber\\
    &+\frac{\tau_0 + \tau_z}{2} \bm h_s \cdot \bm s + \frac{\tau_0-\tau_z}{2} \frac{h_p^3}{\Delta^2}\left[ \cos(3\phi) s_x + \sin (3\phi) s_y \right].
\label{eq:HZeeman}
\end{align}
Here, we  labeled $g_{s}\mu_B\mathbf{B}_{\parallel}\equiv \bm h_s$ and and $g_{p}\mu_B\mathbf{B}_{\parallel}\equiv \bm h_p$.

\subsection{Effect of  $s$-orbital Zeeman Field}
It is one of the main goals of this work to show the importance of including the magnetic field effect on the $p$ orbitals to properly capture features of the low energy bands. As $h_p^3/\Delta^2 \ll h_s$, however, we can first ignore it to find the eigenvalues (up to quadratic order in $k$)
\begin{align}
    E_{\alpha\beta n} &= \epsilon_n( \mathbf{k}) + \alpha \frac{h_s}{2} + \beta \sqrt{A^2 k^2 + \left(  \mathcal M_n(\mathbf{k})  + \alpha\frac{h_s}{2}\right)^2}.
    \label{eq:dispersion_hp_zero}
\end{align}
Here $\beta=\pm$ denotes the $S$ and $P$ orbital character of each band at $\Gamma$, respectively, while $\alpha=\pm$ denotes the two states in the Kramers doublet of a given orbital. 
 There are two distinct regimes that display different behavior at the $\Gamma$ point, where the energies simplify to
\begin{align}
	E_{\alpha\beta n}(\mathbf{k}=0) &= \epsilon_n + \alpha \frac{h_s}{2} + \beta \left |  M_n + \alpha \frac{h_s}{2} \right|
\end{align}
In the weak field regime, when $\left| M_n\right| > \frac{h_s}{2}$, the system is gapped at $\Gamma$. Focusing on the interesting case of a topological gap, $M_n >0$, we find
\begin{align}
	E_{\alpha\beta n}(\mathbf{k}=0) &= \epsilon_n + \alpha (1+\beta)\frac{h_s}{2} + \beta  M_n \,.
\end{align}
The $s$ bands ($\beta = +$) are spin-split, whereas the $p$ bands ($\beta = -$) are degenerate at $\Gamma$. 

In the strong field regime, when $h_s >  2 M_n$, the $\alpha = -$ bands have been un-inverted as the gap for the $\alpha = -$ s-orbital band at $\Gamma$ is given by $M_n - \frac{h_s}{2} < -M_n$. The un-inversion of the $\alpha = -$ bands leads to a quadratic band touching between the $p-$orbital bands. As we will see below, the inclusion of the $p-$orbital Zeeman field resolves the quadratic band touching into two Dirac points perpendicular to the applied field, provided the field points perpendicular to a mirror plane. We estimate that for sample sizes $L = 19$ nm the relevant subband $n=2$ has a topological gap of size $M_{n=2} \approx 5.5$~meV such that the strong field regime is reached by Zeeman fields $h_s \sim 11$ meV. For a $g-$factor of $g_s = 12$~\cite{guoQuantumHallState2022} this corresponds to fields $B_\parallel \sim 16$ T.

\begin{figure}[tb]
\centering
\includegraphics[width=\linewidth]{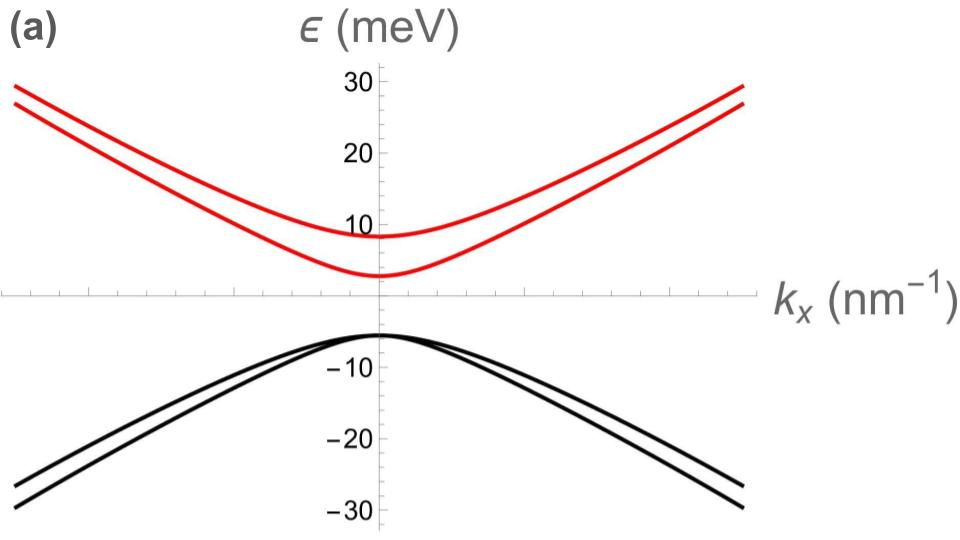}
\includegraphics[width=\linewidth]{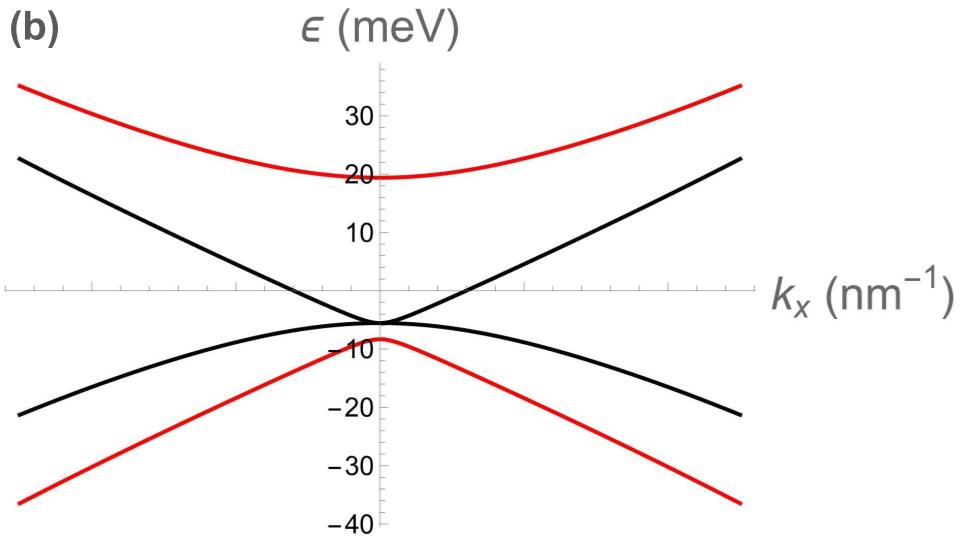}
\label{fig:1}
\caption{Band structure plots for thin film Cd$_3$As$_2$ in the low and high field regimes when setting $h_p = 0$. The dispersion is given by Eq.~\eqref{eq:dispersion_hp_zero} with model coefficients taken from Ref.~\cite{Olle_PRB_2020} (also listed at the end of Appendix~\ref{sec:App_Surface_States}) (a) Low-field dispersion for $h_s =  M_n/2$ with non-degenerate subbands. The bands which arise from the $s-$orbitals at the $\Gamma$ point are colored red, and the ones arising from the $p-$orbitals are colored black.  (b) High-field dispersion for $h_s =5 M_n/2$, where the $\alpha = -$ bands have un-inverted, resulting in a quadratic band touching of the $p-$bands at $\Gamma$ for $h_p=0$.}
\end{figure}

\subsection{Inclusion of the $p-$orbital Zeeman field}
\label{subsec:inclusion_p_orbital_zeeman}
The inclusion of the $p-$orbital field is only relevant in the large field limit where the $\alpha = -$ bands have been un-inverted, and the low energy theory is controlled by the bands with $p$-orbital character near the $\Gamma$ point. Importantly, as $|\mathbf{k}|$ is increased, there is a significant contribution to these bands from the $s$-orbitals. It is the main goal of this section to show that, by including the $p$-orbital field and considering the terms of cubic order in $k$, the low-energy Dirac fermions are gapped out for specific field orientations.

By integrating out the bands at higher energies (See Appendix~\ref{sec:App_Arb_Coeff} for further details), we arrive at an effective Hamiltonian for the lowest energy bands,
\begin{align}
\label{eq:HpEffective}
    \mathcal H_p = -M_n + \left( \begin{array}{cc}
     \mathcal{B}_2 & \mathcal{B}_1\\
    \mathcal{B}_1^*  & -\mathcal{B}_2
    \end{array}\right)
\end{align}
where, up to quintic order, 
\begin{subequations}
\begin{align}
    \label{eq:B1}
    \mathcal{B}_1 &= \biggl(\frac{A^2 k_-^2}{h_s} + \frac{2A A_1 k^2 k_+^2}{h_s}\biggr) e^{-i\phi} \nonumber \\ & \qquad- \frac{\eta^2 k_-^4}{h_s} \left( \frac{n \pi}{L}\right)^2e^{i\phi}  + \frac{h_p^3}{\Delta^2}e^{-3i\phi}  \\
    \label{eq:B2}
    \mathcal{B}_2 &= - \frac{2\eta  k^3}{h_s}\left(\frac{n \pi}{L}\right)\left[ A \cos(\theta - \phi) + A_1 k^2 \cos(5\theta - \phi) \right]
\end{align}
\end{subequations}
Here, $\theta$ and $\phi$ are defined via $\mathbf{k} = k (\cos \theta, \sin \theta, 0)$ and $\mathbf{B} = B (\cos \phi, \sin \phi, 0)$, respectively, and we have defined $k_- = k_x - i k_y$.
We first consider the eigenvalues up to $\mathcal O (k^2)$, which are given by
\begin{align}
    E_{p\pm} &= - M_n \pm \left |\frac{A^2 k^2}{h_s}e^{-2i(\theta-\phi)}+\frac{h_p^3}{\Delta^2}\right|.
\end{align}
We can see that, to $\mathcal O(k^2)$, two Dirac points are created perpendicular to the applied field direction at $\theta_{c\sigma} = \phi + \sigma \pi/2$ and $k_{c0} = \sqrt{h_s h_p^3}/(\Delta A)$,
where $\sigma = \pm$ denotes the two Dirac points. 

The inclusion of terms $\mathcal O(k^3)$ lowers the artificial symmetry $C_{\infty,z}$ down to $C_{4z}$, and the Dirac fermions become gapped for field directions which are not perpendicular to a high symmetry plane~\cite{Burkov_PRB_2018}. Further, $k_{c0}$ is shifted slightly to $k_c$, which is the solution of $|\mathcal B_1 |= 0$ for $\theta = \theta_{c\sigma}$. This shift, however, is small due to the smallness of $h_p^3/\Delta^2$ in comparison to $h_s$ and therefore $k_c$ is well approximated by $k_{c0}$. (See Appendix~\ref{sec:App_Arb_Coeff} for more details)
We may model the Dirac fermions by expanding $\mathcal B_{1,2}$ about $\mathbf{k}_c = k_c (\cos \theta_{c\sigma}, \sin \theta_{c\sigma})$, keeping the lowest order non-zero contributions in $k_c$, to get
\begin{align}
\label{eq:Hsigma}
    \mathcal H_\sigma &= - M_n + \left(\begin{array}{cc}
        \sigma m(\phi) & \frac{-2iA^2 k_c}{h_s}\delta k_- e^{-2i\phi}\\
        \frac{2iA^2 k_c}{h_s}\delta k_+ e^{2i\phi} & -\sigma m(\phi)
    \end{array}\right)\,,
\end{align}
where $m(\phi) = m_0 \sin(4\phi)$, and $m_0 = \frac{2\eta A_1 k^5_c}{h_s} \left(\frac{n \pi}{L}\right)$ is the mass of the Dirac fermion. We also define $\delta k_\pm = \delta k_x \pm i \delta k_y$ with $\delta k_{x,y}$ being the deviation of the momentum away from $\bm k_c$. Importantly, the mass $m(\phi)$ is zero when the field is applied perpendicular to a mirror plane of the relevant point group $D_{4h}$, as derived in Ref.~\cite{Burkov_PRB_2018}. Since the system obeys $D_{4h}$ point symmetry, this corresponds to the field being applied along the $x$ or the $y$ axis as well as along the lines $x= \pm y$.
As the field is rotated through a plane where $m(\phi)$ vanishes, the mass changes sign, switching the chirality of the two flavors of Dirac fermions ($\sigma = \pm$), but keeping the total Chern number of the system unchanged.

\begin{figure}[tb]
\centering
\includegraphics[width=\linewidth]{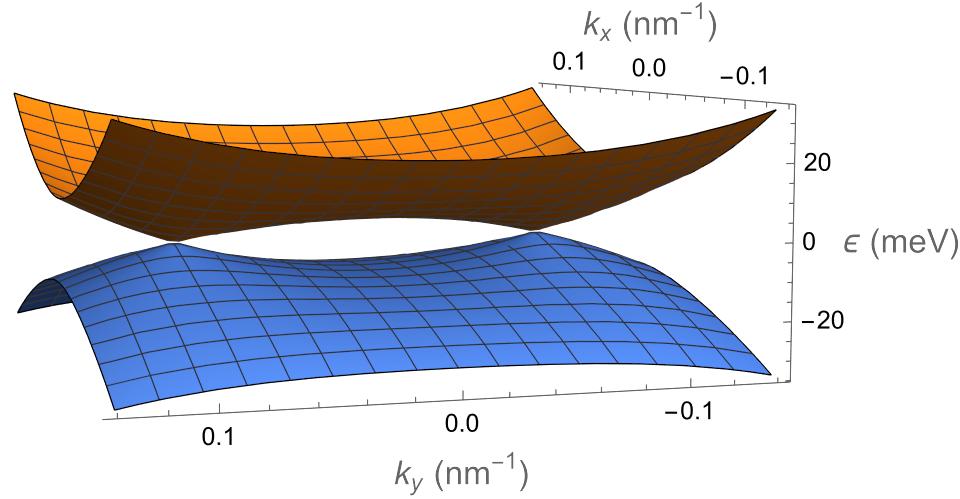}
\caption{3D Plot of the bands from Eq.~\eqref{eq:HpEffective} when $\phi = 0$, showing the two band touchings which give rise to the emergent Dirac states. We use $h_s = 2.5 M_n$, $h_p^3/\Delta^2 = h_s/20$, $A_1 = \eta = 100$ meV$\cdot$nm$^3$, and other model
parameters are taken from Ref.~\cite{Olle_PRB_2020}.
}
\label{fig:POrbitalBands}
\end{figure}

\subsection{Anomalous Hall Effect for tilted field\label{subsec:Anomalous-Hall}}
\begin{figure}[tbh]
\centering\includegraphics[width=.85\linewidth]{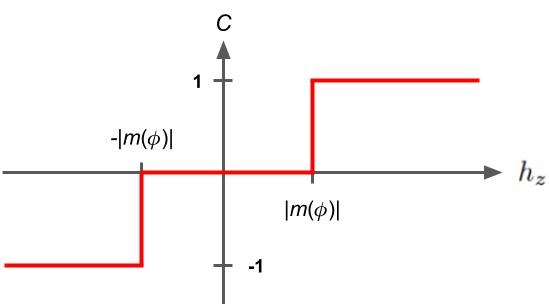}
\caption{The Chern number of the system as a function of out-of-plane Zeeman field $h_z$. When the applied field becomes larger than the mass of each Dirac point, $|h_z| > |m|$, the gap of one of the Dirac points changes sign, leading to a change in the Chern number of the system.}
\label{fig:2}
\end{figure}

We now discuss the effect of tilting the Zeeman field out of the $xy$-plane such that it acquires a finite $B_z$ component. As the basis states are $\alpha = \pm$ states an out-of-plane Zeeman field enters in the form $h_z = 2 g_p \mu_B B_z s_z$ (see also Appendix~\ref{sec:App_Adding-a-magnetic-field}). The Dirac Hamiltonian
becomes
\begin{align}
    \mathcal H_\sigma &= - M_n +  \left(\begin{array}{cc}
        \sigma m(\phi) +  h_z& \frac{-2iA^2 k_c}{h_s}\delta k_- e^{-2i\phi}\\
        \frac{2iA^2 k_c}{h_s}\delta k_+ e^{2i\phi} & - \sigma m(\phi) - h_z
    \end{array}\right).
\end{align}
For sufficiently large $h_z$ such that $|h_z| > |m(\phi)|$, the gap of one Dirac fermion can change sign, leading to a change of the Chern number of the system by $\pm 1$ and causing an anomalous Hall effect. The Chern number of the system as a function of $h_z$ is shown in Fig.~\ref{fig:2}.

\section{Destruction of helical Surface States by in-plane field\label{sec:Surface-states}}
We now return to the case of purely in-plane fields, neglecting the orbital magnetic field, and investigate the evolution of the non-trivial helical edge states in the presence of an in-plane Zeeman field.
Without any applied magnetic field, the effective model for thin film Cd$_3$As$_2$ is that of a QSH insulator, and we, therefore, expect non-trivial helical edge states to arise due to the inverted subbands. After the in-plane field is turned on, we show that the helical edge states become coupled, resulting in a gapping of the surface states originating for each inverted subband. Because for reasonable field strengths $h_p^3/\Delta^2 \ll h_s$, we will neglect the $p$-orbital Zeeman field in the following. 

\subsection{Derivation of surface state solutions}
We impose open boundary conditions in the $y$ direction and consider a sample that is a semi-infinite plane for $y \geq 0$. We thus replace $k_y \rightarrow - i \partial_y$ in Eq.~\eqref{eq:HZeeman}. For simplicity, we only keep terms up to $\mathcal O(k^2)$ and neglect the un-inverted band structure $\epsilon_n(\mathbf{k})$, as including these terms would not alter our conclusions. We then use the ansatz 
\begin{equation}
    \Psi_{\alpha \beta n}(k_x, \lambda_n, y) = \psi_{\alpha \beta}(k_x, \lambda_n) e^{\lambda_n y}\,,
\end{equation}
where $\psi_{\alpha \beta}(k_x, \lambda)$ is a four-spinor, and $\alpha, \beta, n$ denote spin, orbital, and subband indices. Since the sample is located at $y \geq 0$, we look for solutions with $\lambda_n < 0$. The eigenenergies and eigenvectors as a function of $k_x$ and $\lambda$ are found as
\begin{subequations}
\begin{align}
\label{eq:eigenenergies}
 E_{\alpha\beta n}(k_x, - i\lambda)&=\alpha\frac{h_{s}}{2}+\beta\sqrt{A^{2} k^{2}+\Bigl(\mathcal{M}_{n}(\bm{k})+\alpha\frac{h_{s}}{2}\Bigr)^{2}}\\
\label{eq:eigenvectors}
    \psi_{\alpha\beta} (k_x, - i\lambda) &= \mathcal N_{\alpha\beta}(\bm{k})
    \begin{pmatrix}
        -\alpha e^{-i\phi} \frac{E_{\alpha\beta n}(\mathbf k)  +  \mathcal M_n(\mathbf k)}{A k_+}\\
        - \alpha e^{-i\phi} \frac{k_-}{k_+}\\
        - \frac{E_{\alpha\beta n}(\mathbf k)  + \mathcal M_n(\mathbf k)}{A k_+}\\
        1
        \end{pmatrix}\,.
\end{align}
\end{subequations}
Here, $k^2 = k_x^2 - \lambda^2$, $k_\pm = k_x \pm \lambda$, and  $\mathcal N_{\alpha\beta}(\mathbf{k})$ is a normalization factor.
We can express $\lambda$ as a function of energy by solving $E_{\alpha\beta n}(k_x, - i \lambda) = \epsilon$ and obtain four solutions
\begin{align}
    \lambda_{\alpha\zeta}(\epsilon) &= -\sqrt{ k_x^2 +\frac{\tilde B_\alpha}{2} +\zeta \sqrt{\tilde B_\alpha^2/4 -\tilde C_\alpha(\epsilon)}} 
\end{align}
where $\alpha,\zeta = \pm$ and

\begin{subequations}
\begin{align}
    \tilde B_\alpha &= \frac{A^2 - 2 M_{xy} \left(M_n+ \alpha \frac{h_s}{2}\right)}{M_{xy}^2},\\
    \tilde C_\alpha (\epsilon) &= \frac{\left(M_n+ \alpha \frac{h_s}{2}\right)^2 - \left( \epsilon - \alpha \frac{h_s}{2}\right)^2}{ M_{xy}^2}.
\end{align}
\end{subequations}
The wavefunction is then
\begin{align}
    \Psi(k_x,y) &= \sum_{\alpha\zeta\beta} \tilde{\mathcal{N}}_{\alpha\zeta\beta} \psi_{\alpha\beta}[k_x, -i\lambda_{\alpha\zeta}(\epsilon)] e^{\lambda_{\alpha\zeta}(\epsilon) y}
\end{align}
where $\tilde{\mathcal{N}}_{\alpha\zeta\beta}$ are constants and the spinor $\psi_{\alpha\beta} (k_x,-i\lambda_{\alpha \zeta})$ is the eigenvector associated with $E_{\alpha\beta n}$ and given by Eq.~\eqref{eq:eigenvectors}.
The boundary condition is $\Psi (k_x,\lambda_n,y= 0) = 0$, \emph{i.e.} the wavefunction vanishes on the surface. Without loss of generality we set $\tilde{\mathcal{N}}_{\alpha\zeta\beta} =\tilde{\mathcal{N}}_{\alpha\zeta} \delta_{\beta+}$, as for a given $\alpha$ the two $\beta = \pm$ eigenstates give the same solution.
As shown in detail in Appendix~\ref{sec:App_Surface_States}, the boundary conditions can be expressed as a determinant of some matrix $\mathcal{A}$ via $\det \mathcal{A} = 0$, where the columns of $\mathcal{A}$ contain the spinors $\psi_{\alpha +}$.
We then solve the boundary conditions for $\epsilon$ as a function of $k_x$, which are plotted at various in-plane field values to yield the dispersions shown in Figs.~\ref{fig:SurfaceStatesNoField} and Fig.~\ref{fig:SurfaceStatesLargeField}.

\subsection{Zero field results}
\begin{figure}[tb]
\centering
\includegraphics[width=\linewidth]{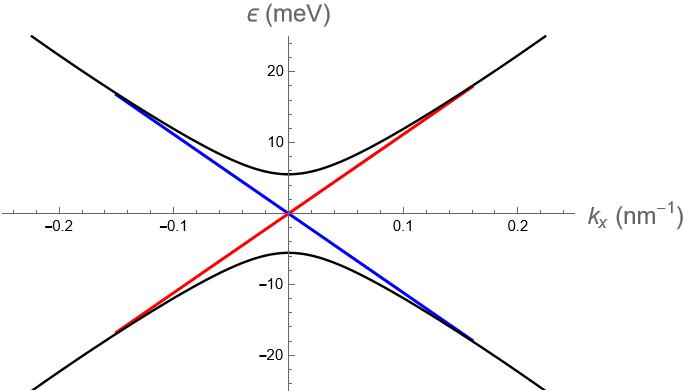}
\caption{Gapless helical surface states on a single Cd$_3$As$_2$ thin film surface in the absence of an in-plane Zeeman field. The $\alpha = +$ surface state is colored red, and the $\alpha = -$ surface state is colored blue, with the bulk bands colored black. Model parameters are taken from Ref.~\cite{Olle_PRB_2020}. We use a film thickness of $L = 19$~nm and show the $n=2$ bulk subbands with energies $\pm M_{n=2} = \pm 5.5$~meV at $\mathbf{k} = 0$.}
\label{fig:SurfaceStatesNoField}
\end{figure}

In the absence of an in-plane Zeeman field, $h_s = 0$, we observe linearly dispersing helical edge states, as expected for a QSH insulator~\cite{ZhouNiu_PRL_2008}. Because the $\alpha = \pm$ subspaces are degenerate, we may simplify our boundary conditions for $\alpha = +$ to
\begin{align}
\label{eq:simplifiedBC}
    \frac{E_{+\beta n}(k_x, -i\lambda_{++})+ \mathcal M_n(k_x, -i\lambda_{++})}{E_{+\beta n}(k_x, -i\lambda_{+-}) + \mathcal M_n(k_x, -i\lambda_{+ -})} = \frac{k_x + \lambda_{+ +}}{k_x+ \lambda_{+-}}
\end{align}
which we solve numerically for the energy $\epsilon$ as a function of $k_x$. Note that the energy $\epsilon$ enters into Eq.~\eqref{eq:simplifiedBC} quadratically, and the correct root is chosen such that the $\beta$ surface state connects to the $\beta$ bulk band. The $\alpha = -$ surface state can then be obtained via the application of time reversal.
As seen in
Fig.~\ref{fig:SurfaceStatesNoField}, the $\alpha = +$ surface state exhibits $v_x > 0$ and the $\alpha = -$ surface state exhibits $v_x < 0$, .

It should be noted that for the surface states to exist one must have
\begin{align}
    \frac{A^2}{M_{xy}^2} > 4\frac{M_n}{M_{xy}} >0.
\end{align}
in order for $\lambda_{\alpha\zeta}$ to have a real component. The coefficients used for the numerical evaluation of the surface state dispersion are the same as used in Ref.~\cite{Olle_PRB_2020} (See Appendix~\ref{sec:App_Surface_States} for details).

\subsection{Gapping of surface states for nonzero in-plane fields}
\begin{figure}[tb]
\centering
\includegraphics[width=\linewidth]{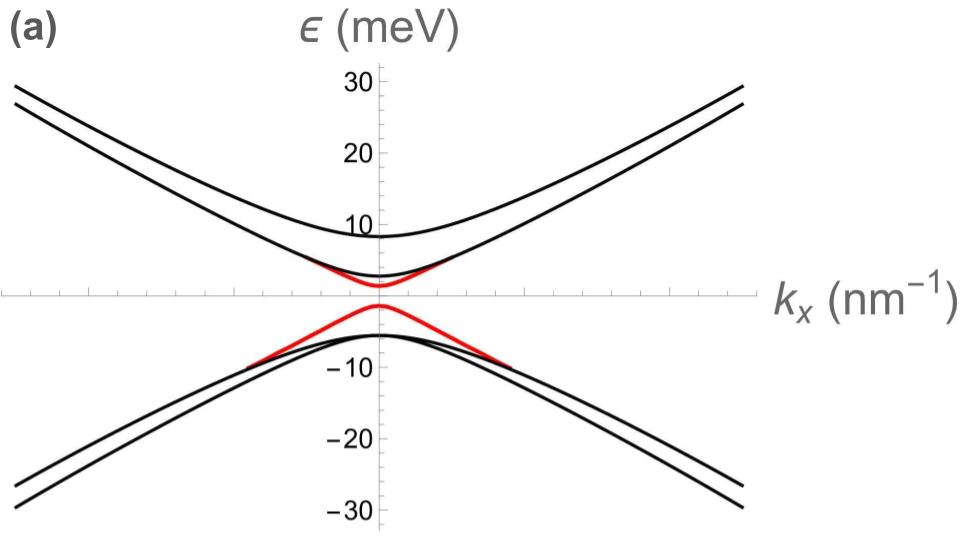}
    \includegraphics[width=\linewidth]{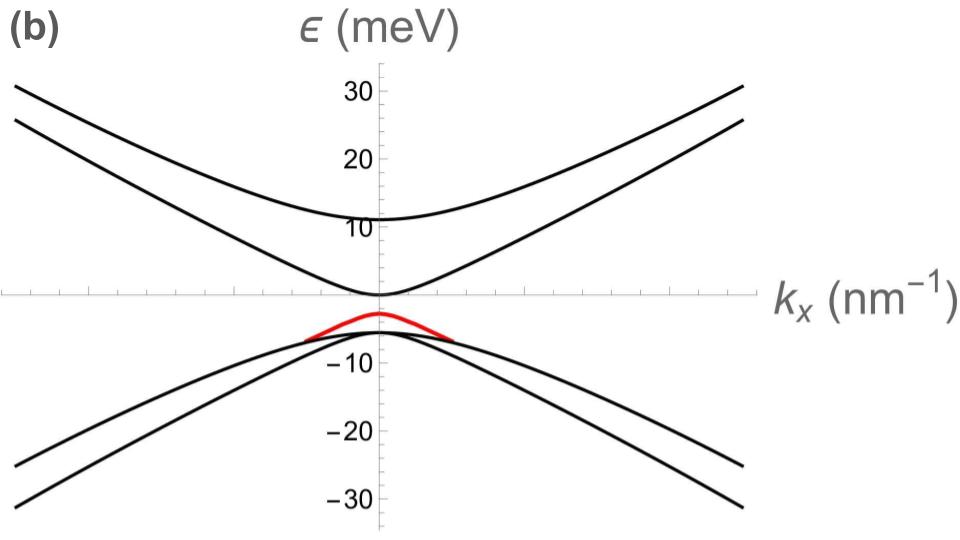} 
\caption{Gapped surface states on a single surface of thin film Cd$_3$As$_2$ in the presence of a small in-plane magnetic field, $h_s < 2 M_n$. The surface states are shown in red and the bulk bands in black. (a) The surface states and bulk bands for $h_s = M_n/2$. (b) The surface state at $h_s= M_n$. where the upper surface state has already merged with the bulk leaving only the lower surface state solution present. Both panels use the model parameters from Ref.~\cite{Olle_PRB_2020}. }
\label{fig:SurfaceStatesLargeField}
\end{figure}

When the Zeeman field is small, $h_s < 2 M_n$, the bulk bands are still inverted, but the helical edge states are now coupled by $h_s$. This opens up a gap in the edge state spectrum on the order of $h_s$, which is consistent with the explicit breaking of time-reversal symmetry. For $h_s < \frac{2}{3} M_n$ there remain two (gapped) surface state solutions (See Fig.~\ref{fig:SurfaceStatesLargeField}\hyperlink{fig:SurfaceStatesLargeField}{(a)}. As the field becomes larger, $h_s \geq \frac{2}{3} M_n$, the $S$-orbital $\alpha = -$ band is pushed through the upper surface state which merges with the bulk. This leaves only a single gapped surface state solution in a small region of momentum space, as shown in Fig.~\ref{fig:SurfaceStatesLargeField}\hyperlink{fig:SurfaceStatesLargeField}{(b)}. 

When the field is increased to $h = 2M_n$, the remaining surface state merges with the bulk bands, and the $\alpha = -$ bulk bands un-invert, and for fields $h_s > 2 M_n$, no surface state solutions arise.

\section{Orbital effect of Out-Of-Plane Magnetic Field \label{sec:Out-plane-B}}
We now include the orbital effects of an out-of-plane magnetic field field, which reorganizes bands into Landau levels. Experimentally, a crossing of the zeroth particle and hole Landau levels as a function of thickness at an out-of-plane field value $B_c$ has been reported~\cite{Stemmer_PRLa_2023}. In subsection \ref{sec:out-plane-field}, we use our model to calculate $B_c$ in terms of k.p parameters, and in subsection \ref{sec:tilted-field} we consider the case of a large in-plane Zeeman field and an out-of-plane orbital field and calculate the Landau level spectrum. 
\subsection{Strictly out-of-plane field results~\label{sec:out-plane-field}}
We first consider a purely out-of-plane field $\mathbf{B} = B \hat{\mathbf{z}}$. Using the Hamiltonian in Eq.~\eqref{eq:H0-third-order} to order $\mathcal O(k^2)$, we introduce a magnetic field via the Peierls substitution,  
$\mathbf{k} \rightarrow -i\bm\nabla + e \mathbf{A}$. We work in the Landau gauge $\mathbf{A} = B y \hat{\mathbf{x}}$.
We then define raising and lowering operators in the conventional way
\begin{subequations}
\begin{align}
    a^\dagger = \frac{l_B}{\sqrt{2}}\left(k_x - \nabla_y + \frac{y}{l_B^2}\right),\\
    a = \frac{l_B}{\sqrt{2}}\left(k_x + \nabla_y + \frac{y}{l_B^2}\right),
\end{align}
\end{subequations}
where $l_B^2 = 1/(|e|B)$ with electron charge $e$.
The Hamiltonian in Eq.~\eqref{eq:H0-third-order}, up to quadratic order in k,   takes the form
\begin{align}
    H &= \left[ M_n - \omega \left(a^\dagger a + \frac{1}{2}\right)\right]\tau_z s_0 \nonumber\\
    &+ \frac{A}{\sqrt{2}l_B}\left[ \tau_x s_z \left(a + a^\dagger\right) + i\tau_y s_0 \left(a-a^\dagger\right) \right], \label{eq:H_a_ops}
\end{align}
where $\omega = 2 M_{xy}/l_B^2$ and $M_n$ is defined in Eq.~\eqref{eq:mass_subband_n}.
Here, we have neglected the band-diagonal contribution $\epsilon_n(\mathbf{k})$, which, as shown in Appendix~\ref{sec:App_LL_Asymm} introduces particle-hole asymmetry but does not change the magnetic field value when the particle and hole Landau levels cross.

Using the trial wavefunctions $\psi_{0} = \left(0 , 0, \ket{0}, 0\right)^T$  and $\psi_{0} = \left(0, \ket{0}, 0, 0 \right)^T$, for the zeroth Landau levels, we find the particle and hole zeroth Landau level energies, 
\begin{align}
    E_{0\pm} &= \pm \left| M_n - \frac{\omega}{2} \right|.
\end{align}
For each subband $n$, there is a crossing of the particle and hole zeroth Landau level energies at a critical magnetic field $B_c(n)$, which is a function of the thickness of the sample,
\begin{align}
    B_c (n) = \frac{1}{|e|M_{xy}} \left[M_0 - M_{z} \left(\frac{n \pi}{L}\right)^2 \right]
\end{align}
For the $n =2$ subbands, which are most relevant for experimental sample thicknesses of $18-20$ nm reported in Refs.~\cite{Stemmer_PRLa_2023,Stemmer_PRLb_2023} and using realistic $k\cdot p$ model parameters from Ref.~\cite{Olle_PRB_2020} (see also Appendix~\ref{sec:App_Surface_States}), we find the critical field $B_c(n=2) \approx 14$~T for $L=18$~nm thick samples. For a slightly larger thickness $L=20$~nm, we find $B_c(n=2) \approx 38$~T. Experiment~\cite{Stemmer_PRLa_2023} reports a critical magnetic field of $\approx 10$~T for $20$~nm thick films, in reasonable agreement with the model estimate.

\subsection{Tilted field results}
\label{sec:tilted-field}
In the presence of a strong in-plane Zeeman coupling ($h_s > 2 M_n$), we have shown previously that the low-energy physics is controlled by the bands arising from the $p$-orbitals at $\Gamma$. The $s$-orbital bands at the $\Gamma$ point are split off in energy by the large Zeeman field $h_s > 2 M_n$. We thus include an out-of-plane orbital field to the effective Hamiltonian $\mathcal H_p$ in Eq.~\eqref{eq:HpEffective}, in which we keep only $\mathcal O(k^2)$,  to find
\begin{align}
\label{eq:Hporb}
     \mathcal H_p &= \left( \begin{array}{cc}
        -M_n + \omega(a^\dagger a + \frac{1}{2}) & \alpha \left(a^\dagger\right)^2+ \frac{h_p^3}{\Delta^2} e^{-3i\phi}\\
        \alpha^* a^2 + \frac{h_p^3}{\Delta^2}e^{3i\phi} & -M_n + \omega(a^\dagger a + \frac{1}{2}) 
    \end{array}\right) 
\end{align}
where $\alpha =  2A^2 e^{-i\phi}/(l_B^2 h_s)$.
In the following, we assume that the in-plane field is oriented along the $x$-axis and set $\phi=0$ unless otherwise stated. Since this direction is perpendicular to a mirror plane, there exist emergent gapless Dirac states 
at $\mathbf{k}_c = \sigma k_c \hat{\mathbf{y}}$ as described in Sec.~\ref{subsec:inclusion_p_orbital_zeeman}. 
\begin{figure}[tb]
\centering
\includegraphics[width=.6\linewidth]{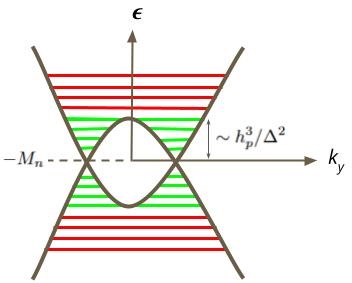}
\caption{Schematic plot of the Landau levels arising from $\mathcal H_p$ in Eq.~\eqref{eq:Hporb} and their transition from being twofold degenerate near the two Dirac points (green) to exhibiting no additional degeneracy (red). The Dirac points occur at $\epsilon = -M_n$ and the transition from double to single degenerate Landau levels occurs at $\epsilon \sim - M_n + h_p^3/\Delta^2$ for the particle Landau levels. }
\label{fig:PotentialSchematic}
\end{figure}

As schematically shown in Fig.~\ref{fig:PotentialSchematic}, there are two distinct regimes: for low-lying Landau levels (shown in green), there is an emergent new quantum number $\sigma$, which denotes the Dirac node and gives rise to a twofold degeneracy of the Landau levels. Above some Landau level index $\nu_c$, the Dirac states are no longer relevant, and this twofold Landau level degeneracy is no longer present.
We consider here only the positive $\nu$ -- the negative $\nu$ can be obtained using particle-hole symmetry of our model (neglecting $\epsilon(\mathbf{k})$). The effect of including the particle-hole symmetry breaking term $\epsilon(\mathbf{k})$ is discussed in detail in Appendix~\ref{sec:App_LL_Asymm}.

We proceed by first finding analytical solutions for the two cases, $\nu > \nu_c$ and $\nu < \nu_c$, and then show numerical results demonstrating the crossover and estimate the crossover index $\nu_c$.

\subsubsection{High energy Landau Levels, $\nu > \nu_c$}
We first consider high energy states and solve Eq.~\eqref{eq:Hporb} for large Landau Level indices $\nu$ such that $\alpha \nu \gg h_p^3/\Delta^2$ and we can neglect the $p-$orbital Zeeman field $h_p$. In this case, the trial wavefunction
\begin{align}
    \psi_\nu &= \left(\begin{array}{c}
        \tilde{C}_{1,\nu} \ket{\nu}\\
        \tilde{C}_{2,\nu} \ket{\nu-2}
    \end{array}\right),
\end{align}
is an eigenstate of $ \mathcal H_p$ [Eq.~\eqref{eq:Hporb}] with eigenvalues
\begin{align}
\label{eq:LLlarge}
    E_{\pm}(\nu)=-M_{n}+\omega\left(\nu-\frac{1}{2}\right)\pm\sqrt{\omega^{2}+|\alpha|^{2}\nu(\nu-1)}.
\end{align}
We see that the eigenvalues are characteristic of Landau levels in a quadratic band touching~\cite{mccannLandauLevelDegeneracyQuantum2006}; schematically, in  Fig.~\ref{fig:PotentialSchematic} they are shown in red.

\subsubsection{Low energy Landau Levels, $\nu < \nu_c$}
At low energies, the physics is dominated by the Dirac states, which occur at $\mathbf{k}_c = \pm \frac{\sqrt{h_p^3h_s}}{A\Delta} \hat{\mathbf{y}}$. Using the expansion of Eq.~\eqref{eq:Hsigma} about these points, $\mathbf{k} = \delta k_x \hat{\mathbf{x}} + (\sigma k_c +\delta k_y) \hat{\mathbf{y}}$, where $\sigma = \pm$ denotes the two band touching points, and inserting the Peierls substitution for $\delta k$, the Hamiltonian is

\begin{align}
    \mathcal H_{\sigma} &= \left( \begin{array}{cc}
        -M_n & -2i\frac{A^2}{h_s}k_c\left(\delta k_x - i \delta k_y\right)\\
        2i \frac{A^2}{h_s} k_c \left( \delta k_x + i \delta k_y\right) & - M_n
    \end{array}\right).
\end{align}
Inserting the orbital field via Peierls substitution for the deviations $\delta \mathbf{k}$, we get the Hamiltonian of each Dirac state
\begin{align}
     \mathcal{H}_\sigma &= \left(\begin{array}{cc}
        -M_n  & \Gamma a^\dagger \\
         \Gamma^* a & -M_n 
    \end{array}\right).
\end{align}
where
\begin{equation}
    \Gamma = -i  \frac{2^{3/2}A^2 k_c}{h_s l_B} e^{-2i\phi}.
\end{equation}
The wavefunction is 
$\psi_{\nu}=\left(-i\ket{\nu},\ket{\nu-1}\right)^{T},$ 
 while the eigenenergies read
\begin{align}
\label{eq:LLsmall}
    E_{\sigma \pm}(\nu) &= -M_n \pm \left| \Gamma\right|\sqrt{|\nu|} \,
\end{align}
and we see that the Landau level spectrum about each Dirac point is identical, yielding a two-fold degeneracy for each Landau level, as indicated by the green Landau levels in Fig.~\ref{fig:PotentialSchematic}.

Notably, if the in-plane field is rotated away from the high-symmetry directions, a finite mass term $m(\phi)$ arises. If we also now include the out-of-plane Zeeman field $h_z$,  the Hamiltonian for the Landau levels due to a single Dirac fermion is given by
\begin{align}
    \mathcal H_\sigma &= -M_n + \left( \begin{array}{cc}
         \sigma m(\phi) + h_z & \Gamma a^\dagger  \\
        \Gamma^* a & -\sigma m(\phi) - h_z
    \end{array}\right)
\end{align}
The wavefunction for the zeroth Landau level is $\psi_0 = (\ket{0}, 0)^T$, giving for the zeroth Landau level energies

\begin{align}
    E_{\sigma}(0) &= - M_n + \sigma m(\phi) + h_z. 
\end{align}
We see that an out-of-plane Zeeman field $h_z$ shifts the zeroth Landau levels of both Dirac fermions in the same direction, as shown in Fig.~\ref{fig:NumericalLLCrossover}. However, the mass $m(\phi)$ term splits the zeroth Landau levels of the two Dirac fermions if the in-plane component of the field does not point in a high symmetry direction. We note that in experiment, Ref.~\cite{Stemmer_PRLb_2023}, the presence of the $\nu = 0$ plateau for large in-plane fields indicates the splitting of the zeroth Landau levels. If the origin of that splitting is the finite $m(\phi)$, then we expect that this plateau should disappear or become harder to see when the in-plane component of magnetic field is pointing in a high-symmetry direction. However, the splitting may also be due to the appearance of correlated, ``emergent" spontaneously valley-polarized states. Future experiment should help to resolve which of the possibilities is realized in practice.

\subsubsection{Crossover regime}
\begin{figure}[tb]
\includegraphics[width=\linewidth]{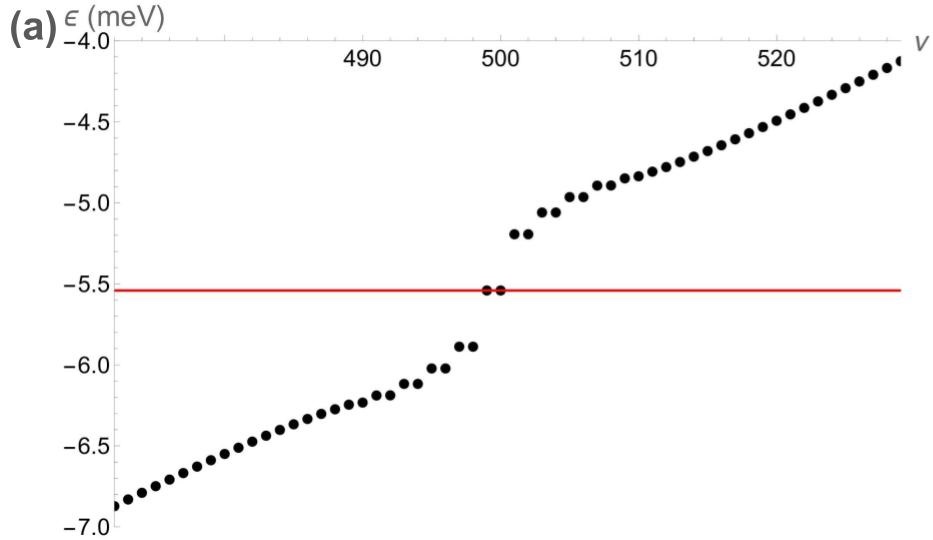}
\includegraphics[width=\linewidth]{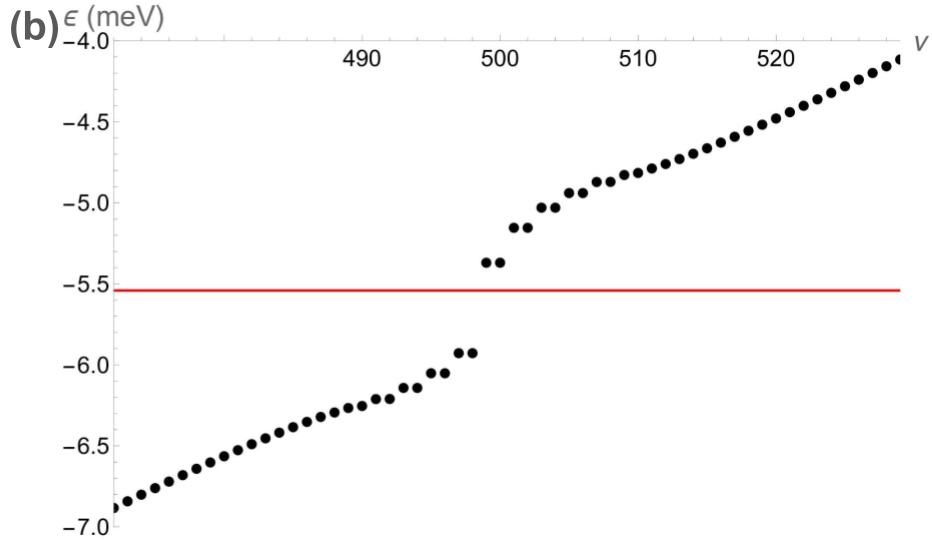}
\caption{$1000$ Landau levels are coupled via Eq.~\eqref{eq:Hporb} and the resulting Hamiltonian is diagonalized numerically. The Landau levels plotted are near $\epsilon = -M_n \approx -5.5$ meV, denoted by the red line. The Landau level energies are calculated at $l_B = 200$~nm, $h_s = 5 M_n/2$, $h_p^3/\Delta^2 =  h_s/20$, and the values of the other coefficients are taken from Ref.~\cite{Olle_PRB_2020}. We have also neglected $\omega$ as it only introduces particle-hole asymmetry.
(a) Landau level spectrum in the absence of $h_z$. The Landau levels are numbered such that $\nu= 501$ occurs at energy $-M_n + |\Gamma|$ (corresponding to $\nu = 1$ in Eq.~\eqref{eq:LLsmall}), and $\nu < 500$ correspond to Landau levels with energies at or below $-M_n$.  (b) Landau level spectrum in the presence of $h_z = 2 g_p B_z$, where we use $g_p = 1/20$ to emphasize the shift of the zeroth Landau levels of the Dirac fermions in the presence of $h_z$.
}
\label{fig:NumericalLLCrossover}
\end{figure}

We have established that the Hamiltonian in Eq.~\eqref{eq:Hporb} exhibits twofold degenerate Landau levels for small $\nu$, one coming from each band touching point at $\mathbf{k} = \sigma k_c \hat{\mathbf{y}}$, with eigenvalues given by Eq.~\eqref{eq:LLsmall}. For large $\nu$, we found singly degenerate Landau levels as one can eventually neglect the $p-$orbital Zeeman field, and the Landau levels lose this twofold degeneracy, with energies given by Eq.~\eqref{eq:LLlarge}. The crossover between these two regimes can be approached numerically by considering a finite number of Landau levels coupling via Eq.~\eqref{eq:Hporb}. Taking a total of $1000$ levels into account, we show the resulting Landau level spectrum in Fig.~\ref{fig:NumericalLLCrossover}. The spectrum is particle-hole symmetric around energy $-M_n$, as we are neglecting the band diagonal $\epsilon_n(\mathbf{k})$ as well as $\omega$, and the index $\nu = 500$ corresponds to the first Landau level above $-M_n$ (lowest green line in Fig.~\ref{fig:PotentialSchematic}). Clearly, the Landau levels exhibit a crossover between being doubly degenerate at low $|\nu-500| < \nu_c$ to being singly degenerate at larger $|\nu-500| > \nu_c$. In the figure, the crossover occurs at $|\nu-500| = 8$. 

The crossover between these two regimes occurs when the Landau level energy is on the order of $E \approx -M_n + \frac{h_p^3}{\Delta^2}$ 
as this is the energy of the $\alpha = -$ $p$-orbital band at the $\Gamma$ point, and illustrated in Fig.~\ref{fig:PotentialSchematic}. 
We can, therefore, estimate the number of Landau levels with a twofold degeneracy, $\nu_c$, using Eq.~\eqref{eq:LLsmall} to obtain
\begin{align}
    \nu_c &\approx 2 \frac{h_p^6}{\Delta^4|\Gamma|^2} = 2 \frac{h_p^3 h_s l_B^2}{2^3 \Delta^2 A^2}\,.
\end{align}
The factor of two in front accounts for the double degeneracy. For the parameters used in Fig.~\ref{fig:NumericalLLCrossover}, we find $\nu_c \approx 7.7$, which is in good agreement with the numerical results. 

Ref.~\cite{Stemmer_PRLb_2023} observes quantum Hall plateaus in a magnetic field with a fixed magnitude $|\bm B| = \sqrt{B_\perp^2 + B_\parallel^2}$ tilted out of the plane at varying angles $\theta_{tilt}$, $\bm B = (B_\parallel \cos\theta_{tilt}, 0 , B_\perp \sin\theta_{tilt})$. 
For smaller tilt angles, which correspond to larger in-plane magnetic fields, the quantum Hall plateaus are observed at odd integer filling factors $\nu= 1,3,5,...$
while for fields with a larger tilt angle quantum Hall plateaus are observed at every integer filling factor.
As the magnetic field magnitude is kept fixed while the field is rotated out of the plane, both $h_s$, $h_p$, and $l_B$ decrease, decreasing $\nu_c$ in agreement with our model. For in plane magnetic field value $B_\parallel = 16$ T, which corresponds to the high field regime, $g_s = 12$,  $h_p^3/\Delta^2 = h_s/20$, and $A =111.6$ meV$\cdot$nm from Ref.~\cite{Olle_PRB_2020},
we calculate that $\nu_c < 2$ for $B_{\perp} < 6.8$ T, in good agreement with experiment which sees the transition between these two regimes occur between $ B_\perp = 3.62$ T and $4.79$ T. 

\section{Conclusions\label{sec:Conclusions}}
In this article, we have derived a low-energy $k \cdot p$ model for $[001]$ thin film Cd$_3$As$_2$ in the presence of a magnetic field applied in an arbitrary direction. At zero field and above a critical thickness, specific subbands $n$ of thin film Cd$_3$As$_2$ are in a quantum spin Hall insulator regime. Importantly, we find that the $p$-orbital Zeeman field enters the Hamiltonian only to cubic order in the applied field, while it is linear for the $s$-orbitals. This distinction is important to understand why the model enters a 2D Dirac semimetal phase in the presence of a sufficiently large in-plane Zeeman field ($h_s > 2 M_n$) that points  perpendicular to a high symmetry mirror plane. The nodal points appear on a mirror plane and become gapped when the applied in-plane field is rotated away from the  high-symmetry directions in plane. 
The fact that the system is initially gapped for small in-plane fields but enters a gapless regime at larger fields (if applied perpendicular to a mirror plane) can explain the experimental observation of a decrease in longitudinal resistance for sufficiently large in-plane fields~\cite{Stemmer_PRLb_2023}. From the realistic model parameters, for sample sizes $L= 19$ nm with an effective $g-$factor of $g_s=12$ we estimate the large field regime to begin for the $n=2$ subbands at $B_\parallel \sim 16$~T. The energy gap follows a $\sin(4 \phi)$ behavior as the field is rotated through the plane, which we predict to be observable when fitting the resistance at finite temperatures to an Arrhenius law. 
Further, we have investigated the fate of the 1D helical edge states of the zero field QSH insulator in the presence of an in-plane magnetic field. Neglecting the small $p$-orbital Zeeman field, we showed how they gap out and evolve in the presence of an in-plane field that couples the $s$-orbitals. While they gap out at infinitesimal fields, consistent with the explicit breaking of time-reversal symmetry, we further showed their complete removal as the in-plane field $h_s$ is increased. 
We further study the inclusion of a perpendicular orbital magnetic field and show that, in the absence of an in-plane Zeeman field, a crossing of the zeroth Landau levels as a function of film thickness occurs. This behavior is consistent with the experimental observations reported in Ref.~\cite{Stemmer_PRLa_2023}.
Lastly, we investigated how the system responds to a perpendicular orbital magnetic field in the 2D Dirac semimetal regime and find that the resulting Landau level spectrum exhibits a twofold degeneracy up to a Landau level index $\nu_c \propto h_p^3 h_s l_B^2/\Delta^2 A^2$, where $A$ sets the Fermi velocity in the bulk model and $\Delta$ is the energy separation of the low-energy manifold from higher bands. The double degeneracy arises from the emergence of a new quantum number associated with the two 2D Dirac nodes that appear at large in-plane fields. The Landau level spectrum crosses over to be singly degenerate at larger energies for $\nu > \nu_c$. We present analytical results for the Landau level spectrum of the two regimes and show numerical results for the crossover.  These results can explain recent quantum Hall measurements presented in Ref.~\cite{Stemmer_PRLb_2023}, which reported an initial jump of the quantum Hall plateaus by two conductance quanta $(e^2/h$) at small out-of-plane fields before returning to plateaus changing by one for larger out-of-plane fields. 

\section{Acknowledgments \label{sec:Acknowledgments}}
We thank S. Stemmer for useful discussions that motivated this work. We also acknowledge discussions with O. Heinonen, J.X. Zhu, and  N. Fokkens.
This work was supported by the Center for the Advancement of Topological Semimetals (CATS), an Energy Frontier Research Center funded by the U.S. Department of Energy (DOE) Office of Science (SC), Office of Basic Energy Sciences (BES), through the Ames National Laboratory under contract DE-AC02-07CH11358. Research at Perimeter Institute is supported in part by the Government of Canada through the Department of Innovation, Science and Economic Development and by the Province of Ontario through the Ministry of Economic Development, Job Creation and Trade.

\appendix

\section{Derivation of the k.p model~\label{sec:App_kp_model}}
In this Appendix, we derive the effective $k \cdot p$ model up to cubic order,
complementing the derivations presented before, up to second order~\citep{Olle_PRB_2020}.
Up to quadratic order, it matches the model of Refs.~\cite{Stemmer_PRLa_2023,Olle_PRB_2020,wang_three-dimensional_2013}. We consider
the space spanned by the states $\left\{ \left|S_{1/2},1/2\right\rangle ,\left|P_{3/2},3/2\right\rangle ,\left|S_{1/2},-1/2\right\rangle ,\left|P_{3/2},-3/2\right\rangle \right\}$. This basis order corresponds to $s_i \otimes \tau_j$ ordering. In this basis, the matrix structure of the symmetry transformations generating all operations in the point group $D_{4h}$ read
\begin{align}
C_{2x} & =-is_{x}\tau_{z},\\
C_{2y} & =-is_{y}\tau_{0},\\
P & =s_{0}\tau_{z},\\
T & =is_{y}\tau_{0}K,\\
C_{4z} & =  \frac{1}{\sqrt{2}}(s_{0}\tau_{z}-i s_{z}\tau_{0})
\end{align}
Here, $P$ denotes spatial inversion, $T$ denotes time-reversal, and $K$ complex conjugation. 
These symmetry operations can be checked as follows:
\begin{itemize}
\item $C_{2x}$ and $C_{2y}$: There are two ways of finding the operator.
The first is to rewrite the states in terms of spherical harmonics.
For instance, $\left|P_{3/2},3/2\right\rangle $ comes from $Y_{1,1}$
combined with a spin half pointing up. Another way is purely in terms
of the total angular momentum. For that, we write down the operator
for $J=1/2$ and for $J=3/2$. For instance, for $J=3/2$, we need
$C_{2x}=\exp\left(-i\frac{2\pi}{2}J_{x}\right)$, where $J_{x}$ is
the $x$ component of the spin $3/2$. We project onto the states
of $M_{J}=\pm3/2$. We repeat the procedure for the $J=1/2$ case. The
final operator is the direct sum of the two. 
\item $P$: The $p$ orbitals are odd under inversion, while the s orbitals
are even;
\item $T$: Time reversal acts the usual way as these four states consist of two Kramers doublets;
\item $C_{4z}$: we can use the total angular momentum operators, $J=1/2$
and $J=3/2$, and compute how they rotate;
\end{itemize}
Finally, the mirror symmetry operations are:
\begin{align}
M_{xz} & = P \circ C_{2y} = -i s_{y}\tau_{z},\\
M_{(x+y)z} &= C_{4z} \circ M_{xz} =  \frac{i}{\sqrt{2}}(-s_{x}\tau_{z}+ s_{y}\tau_{0})  \\
M_{yz} &= C_{4z}^2  \circ M_{xz}=-i s_{x}\tau_{0} \\
M_{(x-y)z} &= C_{4z}^3 \circ M_{xz} =   -\frac{i}{\sqrt{2}}(s_{x}\tau_{z}+ s_{y}\tau_{0}).  
\end{align}
In general, the Hamiltonian can be expressed as
\begin{align}
H & =\sum_{i,j}\sum_{\mathbf{k}}H_{ij}(\mathbf{k})c_{i\mathbf{k}}^{\dagger}c_{j\mathbf{k}}.
\end{align}
We call $U_{g}$ the unitary part of the symmetry, acting on the orbital
and spin indices, and $R_{g}$ the representation of the symmetry
in momentum space. The symmetry constraints impose that 

\begin{align}
U_{g}H\left(R_{g}^{-1}\mathbf{k}\right)U_{g}^{-1} & =H\left(\mathbf{k}\right).
\end{align}
The $k \cdot p$ Hamiltonian satisfying the symmetries can be computed using algebraic subroutines that incorporate all the symmetry operations. We use the package MagneticKP for Mathematica~\cite{ZHANG_magneticKP_2023}. Up to order $k^{3}$, we arrive at Eq.~\eqref{eq:H0-third-order} of the main text.

\section{Microscopic derivation of the Zeeman and spin-orbit terms~\label{sec:App_Adding-a-magnetic-field}}
In this Appendix, we derive the effects of adding a magnetic field along a general direction to thin films of Cd$_{3}$ As$_{2}$. This Appendix is organized in
the following way. First, in Sec.~\ref{subsec:change-basis}
we list the different bases: the spherical basis, the Cartesian one, and
the one that diagonalizes the total angular momentum. In Sec.~\ref{subsec:SOC},
we write down the spin-orbit coupling term on the different bases. In Sec.~\ref{subsec:Zeeman_terms},
we derive the terms in the Hamiltonian corresponding to a magnetic field. 

\subsection{The different basis and how to relate them~\label{subsec:change-basis}}
Throughout this Appendix, we are going to change basis several times.
First, we write down all the operators on the spherical basis, ordered according to

\begin{align}
\left\{ \left|\uparrow\right\rangle ,\left|\downarrow\right\rangle \right\} \otimes\left\{ \left|s\right\rangle ,\left|1,1\right\rangle ,\left|1,0\right\rangle ,\left|1,-1\right\rangle \right\} \nonumber \\
=\left\{ \left|s,\uparrow\right\rangle ,\left|1,1,\uparrow\right\rangle ,\ldots\right\} \,.\label{eq:spherical_basis}
\end{align}
Here, $\ket{l,m}$ are the indices of the spherical harmonics $Y_{lm}$ denoting the orbital angular momentum numbers.
Another useful basis is the one with the $s$ and $p_{x,y,z}$ orbitals,
\begin{equation}
\left\{ \left|\uparrow\right\rangle ,\left|\downarrow\right\rangle \right\} \otimes\left\{ \left|s\right\rangle ,\left|p_{x}\right\rangle ,\left|p_{y}\right\rangle ,\left|p_{z}\right\rangle \right\} =\left\{ \left|s,\uparrow\right\rangle ,\left|p_{x},\uparrow\right\rangle ,\ldots\right\} .\label{eq:pi_basis}
\end{equation}

The basic transformations come from writing down the spherical harmonics
in terms of the $p$ orbitals,
\begin{align}
Y_{1,-1} & =\frac{1}{\sqrt{2}}\left(p_{x}-ip_{y}\right),\label{eq:Y1m1}\\
Y_{1,1} & =\frac{1}{\sqrt{2}}\left(-p_{x}-ip_{y}\right),\label{eq:Y11}\\
Y_{0,0} & =p_{z}.\label{eq:Y00}
\end{align}
Here, we do not explicitly write the spin index; the transformations
are the same for the up and down components. Conversely, 
\begin{align}
p_{x} & =\frac{1}{\sqrt{2}}\left(Y_{1,-1}-Y_{1,1}\right),\label{eq:px}\\
p_{y} & =\frac{i}{\sqrt{2}}\left(Y_{1,-1}+Y_{1,1}\right),\label{eq:py}\\
p_{z} & =Y_{1,0}.\label{eq:pz}
\end{align}
The rotation matrix from the spherical to the $p_{x,y,z}$ basis can
be read off from Eqs.~(\ref{eq:Y1m1}), (\ref{eq:Y11}), and (\ref{eq:Y00}), 

\begin{equation}
U=\left(\begin{array}{cccccccc}
1 & 0 & 0 & 0 & 0 & 0 & 0 & 0\\
0 & -\frac{1}{\sqrt{2}} & \frac{i}{\sqrt{2}} & 0 & 0 & 0 & 0 & 0\\
0 & 0 & 0 & 1 & 0 & 0 & 0 & 0\\
0 & \frac{1}{\sqrt{2}} & \frac{i}{\sqrt{2}} & 0 & 0 & 0 & 0 & 0\\
0 & 0 & 0 & 0 & 1 & 0 & 0 & 0\\
0 & 0 & 0 & 0 & 0 & -\frac{1}{\sqrt{2}} & \frac{i}{\sqrt{2}} & 0\\
0 & 0 & 0 & 0 & 0 & 0 & 0 & 1\\
0 & 0 & 0 & 0 & 0 & \frac{1}{\sqrt{2}} & \frac{i}{\sqrt{2}} & 0
\end{array}\right).\label{eq:U_spher_pi}
\end{equation}
The first $4\times4$ block corresponds to spin up, while the second
one corresponds to spin down. Finally, since spin-orbit coupling is
large, we will work with the total angular momentum basis

\begin{align}
\mathcal{H} & =\left\{ \left|S_{1/2},1/2\right\rangle ,\left|S_{1/2},-1/2\right\rangle ,\left|P_{3/2},3/2\right\rangle ,\left|P_{3/2},1/2\right\rangle ,\right.\nonumber \\
 & \left.\left|P_{3/2},-1/2\right\rangle ,\left|P_{3/2},-3/2\right\rangle ,\left|P_{1/2},1/2\right\rangle ,\left|P_{1/2},-1/2\right\rangle \right\} .\label{eq:tot_ang_mom_basis}
\end{align}
Later, we will consider the low-energy modes, in which we truncate the Hilbert space, $\mathcal{H}_{trunc}$, spanned by the four states (note the $\tau_j \otimes s_i$ ordering here):
\begin{equation}
\left\{ \left|S_{1/2},1/2\right\rangle ,\left|S_{1/2},-1/2\right\rangle ,\left|P_{3/2},3/2\right\rangle ,\left|P_{3/2},-3/2\right\rangle \right\} .\label{eq:hilbert_trunc}
\end{equation}

To change basis to the total angular momentum basis, we have to write down the $J$ multiplets in terms of the original states. The $S_{1/2}$ are trivially
found from the $s$ orbitals,

\begin{align}
\left|S_{1/2},1/2\right\rangle  & =\left|s,\uparrow\right\rangle ,\label{eq:s1/2_up}\\
\left|S_{1/2},-1/2\right\rangle  & =\left|s,\downarrow\right\rangle .
\end{align}
The four $P_{3/2}$ are constructed from the $L=1$, $S=1/2$ states
as 
\begin{align}
\left|P_{3/2},3/2\right\rangle  & =\left|m_{\ell}=1,\uparrow\right\rangle ,\nonumber \\
 & =\frac{1}{\sqrt{2}}\left(-\left|p_{x},\uparrow\right\rangle -i\left|p_{y},\uparrow\right\rangle \right),\\
\left|P_{3/2},1/2\right\rangle  & =\frac{1}{\sqrt{3}}\left(\sqrt{2}\left|m_{\ell}=0,\uparrow\right\rangle +\left|m_{\ell}=1,\downarrow\right\rangle \right),\nonumber \\
 & =\frac{1}{\sqrt{3}}\left(\sqrt{2}\left|p_{z},\uparrow\right\rangle -\frac{1}{\sqrt{2}}\left|p_{x},\downarrow\right\rangle -\frac{i}{\sqrt{2}}\left|p_{y},\downarrow\right\rangle \right),\\
\left|P_{3/2},-1/2\right\rangle  & =\frac{1}{\sqrt{3}}\left(\left|m_{\ell}=-1,\uparrow\right\rangle +\sqrt{2}\left|m_{\ell}=0,\downarrow\right\rangle \right),\nonumber \\
 & =\frac{1}{\sqrt{3}}\left(\frac{1}{\sqrt{2}}\left|p_{x},\uparrow\right\rangle -\frac{i}{\sqrt{2}}\left|p_{y},\uparrow\right\rangle +\sqrt{2}\left|p_{z},\downarrow\right\rangle \right),\\
\left|P_{3/2},-3/2\right\rangle  & =\left|m_{\ell}=-1,\downarrow\right\rangle ,\nonumber \\
 & =\frac{1}{\sqrt{2}}\left(\left|p_{x},\downarrow\right\rangle -i\left|p_{y},\downarrow\right\rangle \right),
\end{align}
Finally, the two $P_{1/2}$ states are also found from the $L=1$, $S=1/2$ as 
\begin{align}
\left|P_{1/2},1/2\right\rangle  & =\frac{1}{\sqrt{3}}\left(\sqrt{2}\left|m_{\ell}=1,\downarrow\right\rangle -\left|m_{\ell}=0,\uparrow\right\rangle \right),\nonumber \\
 & =\frac{1}{\sqrt{3}}\left(-\left|p_{x},\downarrow\right\rangle -i\left|p_{y},\downarrow\right\rangle -\left|p_{z},\uparrow\right\rangle \right),\\
\left|P_{1/2},-1/2\right\rangle  & =\frac{1}{\sqrt{3}}\left(\left|m_{\ell}=0,\downarrow\right\rangle -\sqrt{2}\left|m_{\ell}=-1,\uparrow\right\rangle \right),\nonumber \\
 & =\frac{1}{\sqrt{3}}\left(\left|p_{z},\downarrow\right\rangle -\left|p_{x},\uparrow\right\rangle +i\left|p_{y},\uparrow\right\rangle \right).\label{eq:p1/2_-1/2}
\end{align}
The matrix that changes the basis from $s,p_{x,y,z}$ to the total
angular momentum is immediately found by collecting the results from
Eqs.~(\ref{eq:s1/2_up}) to (\ref{eq:p1/2_-1/2}),

\begin{equation}
U_{J}=\left(\begin{array}{cccccccc}
1 & 0 & 0 & 0 & 0 & 0 & 0 & 0\\
0 & 0 & -\frac{1}{\sqrt{2}} & 0 & \frac{1}{\sqrt{6}} & 0 & 0 & -\frac{1}{\sqrt{3}}\\
0 & 0 & -\frac{i}{\sqrt{2}} & 0 & -\frac{i}{\sqrt{6}} & 0 & 0 & \frac{i}{\sqrt{3}}\\
0 & 0 & 0 & \sqrt{\frac{2}{3}} & 0 & 0 & -\frac{1}{\sqrt{3}} & 0\\
0 & 1 & 0 & 0 & 0 & 0 & 0 & 0\\
0 & 0 & 0 & -\frac{1}{\sqrt{6}} & 0 & \frac{1}{\sqrt{2}} & -\frac{1}{\sqrt{3}} & 0\\
0 & 0 & 0 & -\frac{i}{\sqrt{6}} & 0 & -\frac{i}{\sqrt{2}} & -\frac{i}{\sqrt{3}} & 0\\
0 & 0 & 0 & 0 & \sqrt{\frac{2}{3}} & 0 & 0 & \frac{1}{\sqrt{3}}
\end{array}\right).\label{eq:U_totJ}
\end{equation}
Having all the transformation matrices, we are ready to address how
the Zeeman and the the spin-orbit terms are written in each basis.

\subsection{The spin-orbit coupling~\label{subsec:SOC}}

Before writing down the magnetic field terms, we present the different
ways of writing down the spin-orbit coupling matrix. The spin-orbit coupling can be easily written on the spherical basis by writing
the operators $\mathbf{S}$ for the spin $1/2$ and $\mathbf{L}$ for the spin $1$. The SOC written in matrix form, with the basis
ordered according to Eq.~\eqref{eq:spherical_basis}, reads

\begin{equation}
H_{SO}=\Delta\mathbf{L}\cdot\mathbf{S}=\Delta\left(\begin{array}{cccccc}
\frac{1}{2} & 0 & 0 & 0 & 0 & 0\\
0 & 0 & 0 & \frac{1}{\sqrt{2}} & 0 & 0\\
0 & 0 & -\frac{1}{2} & 0 & \frac{1}{\sqrt{2}} & 0\\
0 & \frac{1}{\sqrt{2}} & 0 & -\frac{1}{2} & 0 & 0\\
0 & 0 & \frac{1}{\sqrt{2}} & 0 & 0 & 0\\
0 & 0 & 0 & 0 & 0 & \frac{1}{2}
\end{array}\right).\label{eq:HSO_spherical}
\end{equation}
Rotating to the $p_{x,y,z}$ basis, of Eq.~(\ref{eq:pi_basis}),
we find, using the rotation matrix of Eq.~(\ref{eq:pi_basis}),

\begin{equation}
\tilde{H}_{SO}=U^{\dagger}H_{SO}U=\frac{\Delta}{2}\left(\begin{array}{cccccccc}
0 & 0 & 0 & 0 & 0 & 0 & 0 & 0\\
0 & 0 & -i & 0 & 0 & 0 & 0 & 1\\
0 & i & 0 & 0 & 0 & 0 & 0 & -i\\
0 & 0 & 0 & 0 & 0 & -1 & i & 0\\
0 & 0 & 0 & 0 & 0 & 0 & 0 & 0\\
0 & 0 & 0 & -1 & 0 & 0 & i & 0\\
0 & 0 & 0 & -i & 0 & -i & 0 & 0\\
0 & 1 & i & 0 & 0 & 0 & 0 & 0
\end{array}\right).\label{eq:HSO_pi_basis}
\end{equation}
It matches the form found in Ref.~\citep{Olle_PRB_2020}.
Finally, we rotate from the $p_{x,y,z}$ to the total angular momentum basis, 

\begin{align}
H_{SOC}^{\left(J\right)}=U_{J}^{\dagger}\tilde{H}_{SOC}U_{J} & =\Delta\left(\begin{array}{cccccccc}
0 & 0 & 0 & 0 & 0 & 0 & 0 & 0\\
0 & 0 & 0 & 0 & 0 & 0 & 0 & 0\\
0 & 0 & \frac{1}{2} & 0 & 0 & 0 & 0 & 0\\
0 & 0 & 0 & \frac{1}{2} & 0 & 0 & 0 & 0\\
0 & 0 & 0 & 0 & \frac{1}{2} & 0 & 0 & 0\\
0 & 0 & 0 & 0 & 0 & \frac{1}{2} & 0 & 0\\
0 & 0 & 0 & 0 & 0 & 0 & -1 & 0\\
0 & 0 & 0 & 0 & 0 & 0 & 0 & -1
\end{array}\right).\label{eq:HSOC_totalJ}
\end{align}
This matrix is diagonal in the total angular momentum basis. It can also be found simply from $\mathbf{L}\cdot\mathbf{S}=\frac{1}{2}\left(\mathbf{J}^{2}-\mathbf{L}^{2}-\mathbf{S}^{2}\right)$.
For the $J=3/2$ multiplet (entries 3 to 6), coming from the $p$ electrons $\left(l=1\right)$ and $s=1/2$, we get $\mathbf{L}\cdot\mathbf{S}=1/2$.
For the cases of $\left(J,L,S\right)=\left(1/2,1,1/2\right)$, we
find $\mathbf{L}\cdot\mathbf{S}=-1$, while for the states coming from $s$ orbitals, $\left(J,L,S\right)=\left(1/2,0,1/2\right)$,
yielding $\mathbf{L}\cdot\mathbf{S}=0$.

\subsection{The effects of SOC and magnetic fields~\label{subsec:Zeeman_terms}}
The Zeeman term can be easily written in the spherical basis by simply writing $g_i \mu_{B}\mathbf{B}\cdot\left(\mathbf{L}+2\mathbf{S}\right)$ for
all degrees of freedom, following the basis defined in Eq.~(\ref{eq:spherical_basis}). Here, $g_i$ denotes the $g$-factor for a given orbital. In this subsection, for compactness, we will set $\mu_{B}=1$, keeping in mind that all terms that follow are proportional to $\mu_{B}$. The Zeeman term reads 
\begin{widetext}
\begin{align}
 & H_{Z}=g_{s}\mathbf{B}\cdot\left(\mathbf{L}+2\mathbf{S}\right)+g_{p}\mathbf{B}\cdot\left(\mathbf{L}+2\mathbf{S}\right) \nonumber \\
 & = \!
 \begin{pmatrix}     
B_{z}g_{s} & 0 & 0 & 0 & g_{s}(B_{x}-iB_{y}) & 0 & 0 & 0\\
0 & 2B_{z}g_{p} & \frac{g_{p}(B_{x}-iB_{y})}{\sqrt{2}} & 0 & 0 & g_{p}(B_{x}-iB_{y}) & 0 & 0\\
0 & \frac{g_{p}(B_{x}+iB_{y})}{\sqrt{2}} & B_{z}g_{p} & \frac{g_{p}(B_{x}-iB_{y})}{\sqrt{2}} & 0 & 0 & g_{p}(B_{x}-iB_{y}) & 0\\
0 & 0 & \frac{g_{p}(B_{x}+iB_{y})}{\sqrt{2}} & 0 & 0 & 0 & 0 & g_{p}(B_{x}-iB_{y})\\
g_{s}(B_{x}+iB_{y}) & 0 & 0 & 0 & -B_{z}g_{s} & 0 & 0 & 0\\
0 & g_{p}(B_{x}+iB_{y}) & 0 & 0 & 0 & 0 & \frac{g_{p}(B_{x}-iB_{y})}{\sqrt{2}} & 0\\
0 & 0 & g_{p}(B_{x}+iB_{y}) & 0 & 0 & \frac{g_{p}(B_{x}+iB_{y})}{\sqrt{2}} & -B_{z}g_{p} & \frac{g_{p}(B_{x}-iB_{y})}{\sqrt{2}}\\
0 & 0 & 0 & g_{p}(B_{x}+iB_{y}) & 0 & 0 & \frac{g_{p}(B_{x}+iB_{y})}{\sqrt{2}} & -2B_{z}g_{p}
 \end{pmatrix}\!.
\label{eq:eq:HZ_spherical}
\end{align}
By rotating to the $p_{x,y,z}$ basis, we find 

\begin{align}
 & \tilde{H}_{Z}=U^{\dagger}H_{Z}U,\nonumber \\
 & =
\begin{pmatrix}
g_{s}B_{z} & 0 & 0 & 0 & g_{s}(B_{x}-iB_{y}) & 0 & 0 & 0\\
0 & g_{p}B_{z} & -ig_{p}B_{z} & ig_{p}B_{y} & 0 & g_{p}(B_{x}-iB_{y}) & 0 & 0\\
0 & ig_{p}B_{z} & g_{p}B_{z} & -ig_{p}B_{x} & 0 & 0 & g_{p}(B_{x}-iB_{y}) & 0\\
0 & -ig_{p}B_{y}g_{p} & ig_{p}B_{x} & g_{p}B_{z} & 0 & 0 & 0 & g_{p}(B_{x}-iB_{y})\\
g_{s}(B_{x}+iB_{y}) & 0 & 0 & 0 & -g_{s}B_{z} & 0 & 0 & 0\\
0 & g_{p}(B_{x}+iB_{y}) & 0 & 0 & 0 & -g_{p}B_{z} & -ig_{p}B_{z} & ig_{p}B_{y}\\
0 & 0 & g_{p}(B_{x}+iB_{y}) & 0 & 0 & ig_{p}B_{z} & -g_{p}B_{z} & -ig_{p}B_{x}\\
0 & 0 & 0 & g_{p}(B_{x}+iB_{y}) & 0 & -ig_{p}B_{y} & ig_{p}B_{x} & -g_{p}B_{z}
\end{pmatrix}.
\label{eq:HZ_pi_basis}
\end{align}
Finally, in the total angular momentum basis, the Zeeman term reads

\begin{align}
 & H_{B}^{\left(J\right)}=U_{J}^{\dagger}H_{Z}U_{J},\nonumber \\
= & \left(\begin{array}{cc}
g_{s}\left(\begin{array}{cc}
B_{z} & B_{x}-iB_{y}\\
B_{x}+iB_{y} & -B_{z}
\end{array}\right) & 0\\
0 & g_{p}\left(\begin{array}{cccccc}
2B_{z} & \frac{2\left(B_{x}-iB_{y}\right)}{\sqrt{3}} & 0 & 0 & \frac{\left(B_{x}-iB_{y}\right)}{\sqrt{6}} & 0\\
\frac{2(B_{x}+iB_{y})}{\sqrt{3}} & \frac{2}{3}B_{z} & \frac{4}{3}\left(B_{x}-iB_{y}\right) & 0 & -\frac{1}{3}\left(\sqrt{2}B_{z}\right) & \frac{\left(B_{x}-iB_{y}\right)}{3\sqrt{2}}\\
0 & \frac{4}{3}\left(B_{x}+iB_{y}\right) & -\frac{2}{3} B_{z} & \frac{2\left(B_{x}-iB_{y}\right)}{\sqrt{3}} & -\frac{\left(B_{x}+iB_{y}\right)}{3\sqrt{2}} & -\frac{1}{3}\left(\sqrt{2}B_{z}\right)\\
0 & 0 & \frac{2\left(B_{x}+iB_{y}\right)}{\sqrt{3}} & -2B_{z} & 0 & -\frac{\left(B_{x}+iB_{y}\right)}{\sqrt{6}}\\
\frac{\left(B_{x}+iB_{y}\right)}{\sqrt{6}} & -\frac{1}{3}\left(\sqrt{2}B_{z}\right) & -\frac{\left(B_{x}-iB_{y}\right)}{3\sqrt{2}} & 0 & \frac{B_{z}}{3} & \frac{1}{3}\left(B_{x}-iB_{y}\right)\\
0 & \frac{\left(B_{x}+iB_{y}\right)}{3\sqrt{2}} & -\frac{1}{3}\left(\sqrt{2}B_{z}\right) & -\frac{\left(B_{x}-iB_{y}\right)}{\sqrt{6}} & \frac{1}{3}\left(B_{x}+iB_{y}\right) & -\frac{B_{z}}{3}
\end{array}\right)
\end{array}\right).\label{eq:HZ_totalJbasis}
\end{align}
\end{widetext}

From the form of the matrix, we can already find many important properties.
Within each subspace of total angular momentum, the field couples
as $g_{J}\mathbf{B}\cdot\mathbf{J}$, with $J$ the total
angular momentum and $g_{J}$ the corresponding Land\'{e} g-factor~\citep{cohen2019quantum}
\begin{equation}
g_{J}=\frac{3}{2}+\frac{S\left(S+1\right)-L\left(L+1\right)}{2J\left(J+1\right)}.
\end{equation}
Within a given angular momentum multiplet, the Zeeman term is proportional to the total angular momentum $J$~\citep{cohen2019quantum}. $J_{x}$ and $J_{y}$ do not couple $M_{J}=3/2$ with $M_{J}=-3/2$. For this multiplet, therefore, only $J_{z}$ will matter (and, therefore, only $B_{z}$). In fact, for the $J=3/2$ states, the Land\'e g-factor is $g_{J=3/2}\left(J=3/2,L=1,S=1/2\right)=4/3$ while for the $J=1/2$ multiplet of $p$ electrons, $g_{J=1/2}\left(J=1/2,L=1,S=1/2\right)=2/3$. The diagonal entries of the matrix are thus $g_{3/2} M_J = \frac43 \times (\pm \frac32) = \pm 2$, $g_{3/2} M_J = \frac43 \times (\pm \frac12) = \pm \frac23$, and $g_{1/2} M_J = \frac23 \times (\pm \frac12) = \pm \frac13$. The $s$ electrons behave purely as spin 1/2 objects, with a Land\'e factor $g_{J=1/2}\left(J=1/2,L=0,S=1/2\right)=2$. The diagonal Zeeman coupling is proportional to $g_{J} M_J$ and thus $g_{1/2} = 2 \times (\pm \frac12) = \pm 1$. As for the $s$ electrons, only the spin is coupled, and $g_{s}s=2\times1/2=1$. 

If we confine ourselves to the low-energy states [Eq.~(\ref{eq:hilbert_trunc})] by simply projecting $H_{B}^{\left(J\right)}$ to those states, we  find $H_{P}=PH_{B}^{\left(J\right)}P$, with a matrix representation

\begin{equation}
H_P=
\begin{pmatrix}
    g_{s}B_{z} & g_{s}(B_{x}-iB_{y}) & 0 & 0\\
g_{s}(B_{x}+iB_{y}) & -g_{s}B_{z} & 0 & 0\\
0 & 0 & 2g_{p}B_{z} & 0\\
0 & 0 & 0 & -2g_{p}B_{z}
\end{pmatrix}.
\label{eq:HZ_projected}
\end{equation}
Here, $P$ is the projector onto the low-energy subspace. We find
that only the $z$ component of the magnetic field couples to the
$J=3/2$, $M_{J}=\pm3/2$ states. This is expected as, within the $J=3/2$
subspace, the Zeeman term is $g_{J}\mathbf{B}\cdot\mathbf{J}$
and the $x$ and $y$ components cannot couple two states with $M_{z}$
differing by $3$. 

\section{Low-energy effective coupling of p-electrons to in-plane fields:
perturbation theory~\label{appendix:EffectiveZeemanCoupling}}

In this Appendix, we consider how the components of the magnetic field parallel to the $xy$ plane couple to the $p$ electrons by performing a perturbative calculation. We can already infer that we need a process that couples the states of $M_{J}=\pm3/2$. Such processes need to change $M_{z}$ by 3 units, which requires a cubic coupling with the magnetic field. We now demonstrate this by an explicit calculation.

We will be focusing on states at the $\Gamma$ point. In what follows, we neglect the effects of crystal fields that preserve the azimuthal quantum numbers and, therefore, do not alter our conclusions. We call $E_{s}$ the energy of the s orbitals and $\Delta_{1}$, $\Delta_{2}$ and $\Delta_{3}$ the three different energies of the different $p$ doublets. Using the basis as defined in Eq.~\eqref{eq:tot_ang_mom_basis}, the Hamiltonian $H_{0}$ is diagonal, with energy $E_{s}$ for the s-orbitals and energies 
$\Delta_{1},\Delta_{2}$ for the $p$-orbitals of $J=3/2$ and azimuthal numbers $\pm 3/2$ and $\pm 1/2$, respectively. The energy of the p-orbitals of $J=1/2$ is $\Delta_{3}$.


These masses are entering to produce differences in the energy of the bands, to zeroth order, and will lead to the $s$ states $\left|S_{1/2},\pm1/2\right\rangle $ and the states $\left|P_{3/2},\pm3/2\right\rangle $ close to the Fermi level. We decompose the Hamiltonian as 
\begin{equation}
H=H_{0}+H_{B}^{\left(J\right)},
\end{equation}
where the perturbation $H_{B}^{\left(J\right)}$ is the term with magnetic field, Eq.~(\ref{eq:HZ_totalJbasis}). We are going to follow the approach of Ref.~\citep{lindgren1974BW}. We define the resolvent operator
\begin{equation}
R=\frac{Q}{E_{0}-H_{0}},
\end{equation}
with $Q=1-P$, where $P$ projects onto the four-dimensional set of states $\{\ket{S_{1/2},\pm1/2}, \ket{P_{3/2},\pm3/2}\}$, and $E_{0}$ the eigenenergies of $H_{0}$. With the projector $P$ and the resolvent $R$, the different corrections of the Hamiltonian can be found in a straightforward way as~\cite{lindgren1974BW}
\begin{align}
\Delta H^{\left(1\right)} & =PH_{B}^{\left(J\right)}P,\\
\Delta H^{\left(2\right)} & =PH_{B}^{\left(J\right)}RH_{B}^{\left(J\right)}P,\\
\Delta H^{\left(3\right)} & =PH_{B}^{\left(J\right)}RH_{B}^{\left(J\right)}RH_{B}^{\left(J\right)}P-PH_{B}^{\left(J\right)}R^{2}H_{B}^{\left(J\right)}PH_{B}^{\left(J\right)}P.
\end{align}
All the terms $\Delta H^{\left(i\right)}$ are effective $4\times4$ models, written in the subspace spanned by the states listed in Eq.~(\ref{eq:hilbert_trunc}). 

The first correction, $\Delta H^{\left(1\right)}$, is equivalent to neglecting the effects of the high-energy bands and leads to the projected Hamiltonian of Eq.~(\ref{eq:HZ_projected}). In what follows, we will consider only the effects on the $P_{3/2}$ states as the higher-order contributions for the $s$ states vanish. 

The term $\Delta H^{\left(2\right)}$ is an isotropic contribution that adds to the diagonal components and reads 
\begin{equation}
\Delta H^{\left(2\right)}=u^{\left(2\right)}\left|\mathbf{B}_{\parallel}\right|^{2}\left(\begin{array}{cc}
1 & 0\\
0 & 1
\end{array}\right).
\end{equation}
Here, $\mathbf{B}_\parallel = (B_x, B_y, 0)$ and 
\begin{equation}
u^{\left(2\right)}=\frac{9\Delta_{1}-\Delta_{2}-8\Delta_{3}}{6\left(\Delta_{1}-\Delta_{2}\right)\left(\Delta_{1}-\Delta_{3}\right)}.
\end{equation}
Finally, the third-order correction reads $\Delta H^{\left(3\right)}=\Delta H_{\perp}^{\left(3\right)}+\Delta H_{\parallel}^{\left(3\right)}$, with

\begin{equation}
\Delta H_{\parallel}^{\left(3\right)}=u_{\parallel}^{\left(3\right)}\left(\begin{array}{cc}
0 & \left(B_{x}-iB_{y}\right)^{3}\\
\left(B_{x}+iB_{y}\right)^{3} & 0
\end{array}\right).
\end{equation} 
and $\Delta H_{\perp}^{\left(3\right)}=u_{\perp}^{\left(3\right)}s_{z}$
Explicitly,
\begin{align}
u_{\parallel}^{\left(3\right)} & =\frac{\left(9\Delta_{1}-\Delta_{2}-8\Delta_{3}\right)\left(3\Delta_{1}+\Delta_{2}-4\Delta_{3}\right)}{18\left(\Delta_{1}-\Delta_{2}\right)^{2}\left(\Delta_{1}-\Delta_{3}\right)^{2}},\\
u_{\perp}^{\left(3\right)} & =\frac{\left(3\Delta_{1}+\Delta_{2}-4\Delta_{3}\right)^{2}}{18\left(\Delta_{1}-\Delta_{2}\right)^{2}\left(\Delta_{1}-\Delta_{3}\right)^{2}}.
\end{align}


Decomposing in terms of the $\mathbf{s}$ and $\boldsymbol{\tau}$ matrices, we find 
\begin{align}
\Delta H^{\left(3\right)} & =\mu_{B}^{3} g_{p}\left(\frac{\tau_{0}-\tau_{z}}{2}\right)\left\{ B_{x}\left(B_{x}^{2}-3B_{y}^{2}\right)s_{x}\right.\nonumber \\
 & \left.-B_{y}\left(B_{y}^{2}-3B_{x}^{2}\right)s_{y}+r\left|\mathbf{B}_{\perp}\right|^{2}B_{z}s_{z}\right\}, 
\end{align}
$r=u_{\perp}^{\left(3\right)}/u_{\parallel}^{\left(3\right)}$ a dimensionless number of order one.

Adding all contributions, we find an effective coupling with the $\mathbf{B}$ field shown in Eq.~\eqref{eq:HB_parallel} of the main text.

\section{Derivation of $\mathcal H_p$ ~\label{sec:App_Arb_Coeff}}
In this Appendix, we explicitly show that the specific relations between the third-order coefficients in the $k \cdot p$ model that we use in the main text to simplify the resulting expressions do not change any of our conclusions. In other words, it is not necessary to make assumptions about the coefficients $C_{3a,b,c,d,e}$ as is done in the main text, and it will be shown that for arbitrary coefficients, the form of the effective Hamiltonian $\mathcal H_p$ in the high field limit takes the same form as in the main text.
To begin, our full Hamiltonian is given by
\begin{align}
\label{eq:GeneralH}
    \mathcal H &= \left( \begin{array}{cccc}
        M_n & A_+ & h_s e^{-i\phi} & F  \\
        A_-  & - M_n & F & \tilde h_p e^{-3i\phi}\\
        h_s e^{i\phi} & F^* & M_n & - A_-\\
        F^* & \tilde h_p e^{3i\phi} & - A_+ &- M_n
    \end{array}\right)
\end{align}
in the basis $(\ket{S, 1/2},\ket{P_{3/2}, 3/2}, \ket{S,-1/2},\ket{P_{3/2},-3/2})$, and we use the shorthand $\tilde h_p = h_p^3/\Delta^2$ and
\begin{subequations}
\begin{align}
    A_\pm &= \tilde A k_\pm \mp iC_{3a}k_x k_y k_\mp + C_{3b} (k_x^3 \pm i k_y^3),\\
    F &= i C_{3e} k_x k_y k_z + C_{3d}k_z (k_x^2 - k_y^2),
\end{align}
\end{subequations}
where we have conveniently written $\tilde A = A - C_{3c}k_z^2$ and use the shorthand $k_z = n\pi/L$.
As the bands with $p$-orbital character at $\Gamma$ have energies $\epsilon \approx - M_n$, we may take the $s$-orbital subspace of the Schr\"odinger equation and solve for the $s$-orbital wavefunctions in terms of the $p$-orbital wavefunctions in the limit $h_s \gg M_n$:
\begin{subequations}
\begin{align}
    s_{1/2} &\approx \frac{- F^* p_{3/2} + A_- p_{-3/2}}{h_s e^{i\phi}}\\
    s_{-1/2} &\approx \frac{- A_+ p_{3/2} - F p_{-3/2}}{h_s e^{-i\phi}}
\end{align}
\end{subequations}
Substituting these solutions back into the equations for the $p$-orbitals we obtain a low energy theory for the relevant bands near $\epsilon = - M_n$, $\mathcal H_p$, with
\begin{subequations}
\begin{align}
    \mathcal{B}_1 &=\frac{A_-^2}{h_s}e^{-i\phi} - \frac{F^2}{h_s}e^{i\phi}+\tilde h_p e^{-3i\phi},\\
    \mathcal{B}_2 &= -\left[ \frac{F^* A_-}{h_s}e^{-i\phi} + \frac{F A_+}{h_s}e^{i\phi} \right].
\end{align}
\end{subequations}
$\mathcal{B}_1$ to $\mathcal O(k^2)$ is given by
\begin{align}
    \mathcal{B}_1^{(0)} &= \frac{\tilde A^2 k_-^2 }{h_s}e^{-i\phi} + \tilde h_p e^{-3i\phi}.
\end{align}
$\mathcal B_1^{(0)} = 0$ occurs at $k_{c0} = \sqrt{\tilde h_p h_s/\tilde A^2}$ and $\theta_{c\sigma} = \phi \pm \pi/2$. Inclusion of higher order in $k$ terms slightly shifts $k_c$  away from $k_{c0}$ and gives rise to a mass term which arises from $\mathcal B_2$, as we will show below. For coefficients in the atomic limit $\mathcal B_1$ is
\begin{align}
    \mathcal B_1 e^{3i\phi} &= \frac{1}{h_s}\left[ A^2 k^2 e^{-2i(\theta - \phi)} + 2A A_1 k^4 e^{2i(\theta + \phi)} \right.\nonumber\\
    &+ \left.A_1^2 k^6 e^{2i(3\theta + \phi)} - \eta^2 k_z^2 k^4 e^{-4i(\theta-\phi)} \right] + \tilde h_p,
\end{align}
the magnitude of which determines $k_c$ when $\theta = \theta_{c\sigma} = \phi + \sigma \pi/2$. Thus, $\mathcal B_1$ evaluated at $\theta = \theta_{c\sigma}$ is 
\begin{align}
\label{eq:B1full}
    \mathcal B_1 e^{3i\phi} &= \frac{-1}{h_s}\left[ A^2 k^2 + 2 A A_1 k^4 \cos4 \phi + A_1^2 k^6 \cos8\phi - \eta^2 k_z^2 k^4\right] \nonumber\\
    &+\tilde h_p - i\frac{2A_1 k^4}{h_s}\left(A + A_1 k^2 \cos 4\phi\right) \sin4 \phi
\end{align}
In general, the equation for $k_c$ using Eq.~\eqref{eq:B1full} is a 6th order polynomial for $k_c^2$. However, as we limit our Hamiltonian Eq.~\eqref{eq:GeneralH} to $\mathcal O(k^3)$, we should only keep up to $\mathcal O(k^4)$ terms in $\mathcal B_1$, and the resulting equation for $k_c$ is determined by
\begin{align}
    & \left[A^2 k_c^2 + 2 A A_1 k_c^4 \cos4 \phi  - \eta^2 k_z^2 k_c^4 - h_s \tilde h_p \right]^2 \nonumber\\
    &+ 4 A_1^2 A^2 k_c^8 \sin^2 4\phi = 0.
\end{align}
which is a quartic equation for $k_c^2$ and thus has a general solution.
Due to the smallness of $\tilde h_p$ compared to $h_s$, however, $k_c$ is well approximated by $k_{c0}$. 
We can expand $\mathcal B_1$ up to lowest order in $k_c$, giving us
\begin{align}
    \mathcal B_1 &\approx -\frac{2i A^2 k_c}{h_s}\delta k_- e^{-2i\phi}.
\end{align}
The general form of $\mathcal{B}_2$ after expressing $\mathbf{k}$ in polar coordinates is
\begin{widetext}
\begin{align}
\label{eq:B2full}
    \mathcal{B}_2 &= -\frac{\tilde A k^3 k_z}{h_s} \left( C_- \cos(\theta - \phi) + C_+ \cos(3\theta + \phi) \right) - \frac{k^5 k_z}{4h_s}\left[\left( (C_{3a}+ C_{3b}) C_{3e}+ 4 C_{3b} C_{3d}\right) \cos(\theta - \phi) \right.\nonumber\\
    &\left.- (C_{3a}-3 C_{3b}) C_+ \cos(3\theta + \phi) + (C_{3a}+C_{3b}) C_- \cos(5\theta - \phi)  \right]
\end{align}
with $C_\pm = C_{3d} \pm C_{3e}/4$. Substituting in the relation between coefficients $C_i$ used in the main text, Eq.~\eqref{eq:B2full} reduces to Eq.~\eqref{eq:B2}. At $\theta = \phi \pm \pi/2$, we have
\begin{align}
    \mathcal{B}_2 = \pm m (\phi)
     = \pm\left[-\frac{\tilde A k_c^3 k_z}{h_s}C_+  \pm\frac{k^5_c k_z}{4 h_s} \left( (C_{3a}- 3 C_{3b})C_+ + (C_{3a}+ C_{3b}) C_- \right) \right]\sin (4\phi)
\end{align}
The mass thus vanishes if the field points perpendicular to a mirror plane, since this mirror then remains a symmetry of the system. Combining time-reversal with the mirror that contains the field direction is also a symmetry.
\section{Surface States Dispersion Calculation~\label{sec:App_Surface_States}}
In this Appendix, we provide details on the calculation of the helical edge modes with and without an in-plane magnetic field. 
The boundary condition for the surface state is expressed in terms of the eigenvectors in Eq.~\eqref{eq:eigenvectors}
\begin{subequations}
\begin{align}
    \psi_{1\alpha}(\bm k_x, \bm k_y) &= -\alpha \mathcal N_{\alpha +}(k_x,k_y) \frac{E_{\alpha +}(k_x, k_y) + M(k_x, k_y)}{Ak_+}\\
    \psi_{2\alpha}(\bm k_x, \bm k_y) &= - \alpha \mathcal N_{\alpha +} e^{-i\phi} \frac{k_-}{k_+}\\
    \psi_{3\alpha}(\bm k_x, \bm k_y) &= -\mathcal N_{\alpha +}(k_x, k_y)\frac{E_{\alpha +}(k_x, k_y) + M(k_x, k_y)}{Ak_+}\\
    \psi_{4\alpha}(\bm k_x, \bm k_y) &= \mathcal N_{\alpha +}(k_x, k_y)
\end{align}
\end{subequations}
the matrix $\mathcal{A}$ whose determinant being zero gives the surface state dispersion $\epsilon(k_x)$ is then given by
\begin{align}
    \mathcal{A} &= \left( \begin{array}{cccc}
        \psi_{1+}(k_x, -i\lambda_{++}) & \psi_{1+}(k_x, -i\lambda_{+-}) & \psi_{1-}(k_x, -i\lambda_{-+}) & \psi_{1-}(k_x, -i\lambda_{--})\\
        \psi_{2+}(k_x, -i\lambda_{++}) & \psi_{2+}(k_x, -i\lambda_{+-}) & \psi_{2-}(k_x, -i\lambda_{-+}) & \psi_{2-}(k_x, -i\lambda_{--})\\
        \psi_{3+}(k_x, -i\lambda_{++}) & \psi_{3+}(k_x, -i\lambda_{+-}) & \psi_{3-}(k_x, -i\lambda_{-+}) & \psi_{3-}(k_x, -i\lambda_{--})\\
        \psi_{4+}(k_x, -i\lambda_{++}) & \psi_{4+}(k_x, -i\lambda_{+-}) & \psi_{4-}(k_x, -i\lambda_{-+}) & \psi_{4-}(k_x, -i\lambda_{--})
        \end{array}\right)
\end{align}
\end{widetext}
We then solve $\det \mathcal{A} = 0$ numerically to obtain the surface state dispersion $\epsilon(k_x)$. We use the parameters from Ref.~\cite{Olle_PRB_2020},
\begin{subequations}
\begin{align}
    A &= 111.6 \,\text{meV} \, \text{nm}\\
    M_0 &= 28.2 \, \text{meV}\\
    M_{z} &= 207.2 \, \text{meV} \, \text{nm}^2\\
    M_{xy} &= 133.2 \, \text{meV} \, \text{nm}^2
\end{align}
\label{eq:kp_parameters}
\end{subequations}
and we use $n = 2$, $L = 19$ nm to model the Cd$_3$As$_2$ thin film setup reported in Refs.~\cite{Stemmer_PRLa_2023, Stemmer_PRLb_2023}. 

\section{Particle-Hole Asymmetry in Landau Level Spectrum~\label{sec:App_LL_Asymm}}
\begin{figure}[h!]
\vspace*{-2.5\baselineskip}
\centering
\includegraphics[width=.9\linewidth]{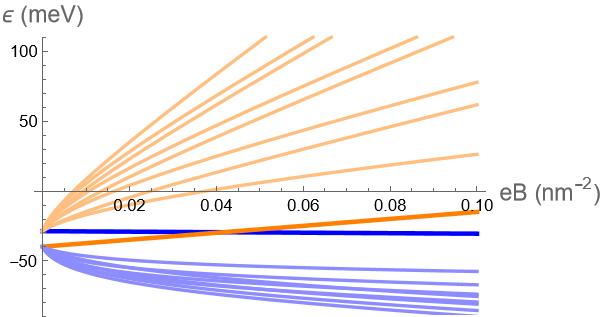}
\caption{
The $\nu = 0-4$ Landau levels plotted as a function of the inverse magnetic length squared $e B$. The $\nu=0$ levels are shown with thicker strokes, with $\nu \neq 0$ particle and hole levels colored orange and blue, respectively. The inclusion of the band-diagonal $\epsilon_n(\mathbf{k})$ breaks particle-hole symmetry by introducing $\omega_2$, which provides a positive dispersion in magnetic field for every Landau level. The values of model coefficients used are from Ref.~\cite{Olle_PRB_2020} [see also Eq.~\eqref{eq:kp_parameters}].}
\label{fig:LandauLevelPHasymmetry}
\end{figure}
The inclusion of the band-diagonal term $\epsilon_n(\mathbf{k})$ [see Eq.~\eqref{eq:eps_n}] causes an additional contribution to Eq.~\eqref{eq:H_a_ops}, and the full Hamiltonian is given by
\begin{align}
    H &= C_n + \omega_2\left (a^\dagger a + \frac{1}{2} \right) +\left[ M_n - \omega \left(a^\dagger a + \frac{1}{2}\right)\right]\tau_z s_0 \nonumber\\
    &+ \frac{A}{\sqrt{2}l_B}\left[ \tau_x s_z \left(a + a^\dagger\right) + i\tau_y s_0 \left(a-a^\dagger\right) \right]
\end{align}
where $C_n = C_0 + C_1\left( \frac{n\pi}{L}\right)^2$ and $\omega_2 = \frac{2C_2 }{l_B^2}$. 
The $\nu \neq 0$ Landau levels are found with the trial wavefunctions
\begin{align}
    \psi_{\nu+} &= \left( 0, 0 ,B_{1+} \ket\nu, B_{2+}\ket{\nu-1}\right)^T\\
    \psi_{\nu-} &= \left( B_{1-} \ket{\nu -1}, B_{2-}\ket\nu, 0 ,0\right)^T
\end{align}
giving the $\nu \neq 0$ Landau level energies
\begin{align}
    \epsilon_{\alpha\beta}(\nu) &= C_n -\alpha \frac{\omega}{2} + \nu\omega_2 \nonumber \\ & \quad + \beta \sqrt{\frac{2 A^2}{l_B^2} \nu + \left( \alpha \left(M_n - \nu \omega \right) + \frac{\omega_2}{2}\right)^2}.
\end{align}
The inclusion of the band-diagonal term causes two contributions to the Landau level spectrum: first, a trivial shift of the overall energies by $C_n$, but more importantly,  $\omega_2$ contributes a positive dispersion to every Landau level as a function of magnetic field, introducing a  particle-hole asymmetry as shown in Fig.~\ref{fig:LandauLevelPHasymmetry}

\end{document}